\documentclass[twocolumn,superscriptaddress,nofootinbib,aps]{revtex4}

\usepackage{amsmath,amssymb}
\usepackage{graphicx}
\usepackage{color}
\usepackage{hyperref}
\usepackage{url}
\usepackage{slashed}
\usepackage{subfigure}
\usepackage{slashed,bbm}
\usepackage{graphics,psfrag,epsfig}
\usepackage{dsfont}
\usepackage{enumerate}

\newcommand{\RNum}[1]{\uppercase\expandafter{\romannumeral #1\relax}}

\newcommand{\rund}[1]{ \left( #1 \right) }
\newcommand{\eck}[1]{ \left[ #1 \right] }

\newcommand{\spitz}[1]{ \left\langle  #1 \right\rangle }
\newcommand{\abs}[1]{ \left|  #1 \right| }
\newcommand{\pr}[1]{{{#1}^{\prime}}}

\begin{document}


\title{Competition of density waves and quantum multicritical behavior in Dirac materials from functional renormalization}

\author{Laura Classen}
\affiliation{Institut f\"ur Theoretische Physik, Universit\"at Heidelberg, Germany}

\author{Igor F. Herbut}
\affiliation{Department of Physics, Simon Fraser University, Burnaby, Canada}

\author{Lukas Janssen}
\affiliation{Department of Physics, Simon Fraser University, Burnaby, Canada}
\affiliation{Institut f\"ur Theoretische Physik, Technische Universit\"at Dresden, Germany}

\author{Michael M. Scherer}
\affiliation{Institut f\"ur Theoretische Physik, Universit\"at Heidelberg, Germany}

\begin{abstract}
We study the competition of spin- and charge-density waves and their quantum multicritical behavior for the semimetal-insulator transitions of low-dimensional Dirac fermions.
Employing the effective Gross-Neveu-Yukawa theory with two order parameters as a model for graphene and a growing number of other two-dimensional Dirac materials allows us to describe the physics near the multicritical point at which the semimetallic and the spin- and charge-density-wave phases meet.
With the help of a functional renormalization group approach, we are able to reveal a complex structure of fixed points, the stability properties of which decisively depend on the number of Dirac fermions $N_f$.
We give estimates for the critical exponents and observe crucial quantitative corrections as compared to the previous first-order $\epsilon$ expansion.
For small $N_f$, the universal behavior near the multicritical point is determined by the chiral Heisenberg universality class supplemented by a decoupled, purely bosonic, Ising sector.
At large $N_f$, a novel fixed point with nontrivial couplings between all sectors becomes stable.
At intermediate $N_f$, including the graphene case ($N_f = 2$) no stable and physically admissible fixed point exists.
Graphene's phase diagram in the vicinity of the intersection between the semimetal, antiferromagnetic and staggered density phases should consequently be governed by a triple point exhibiting first-order transitions.
\end{abstract}

\maketitle


\section{Introduction}

Dirac materials~\cite{wehling2014}, such as graphene~\cite{novoselov2005,geim2007,castroneto2009} and a growing number of novel two-dimensional systems, exhibit a huge variety of possible ordered states in the presence of sufficiently strong interactions~\cite{sorella1992,khvesh2001,gorbar2002,herbut2006,black2007,raghu2008,roy2009,honerkamp2008,grushin2013,daghofer2013,duric,herbut2009}.
%
%
By inreasing, for instance, repulsive onsite or nearest-neighbor density-density interactions, these systems are expected to exhibit a continuous quantum phase transition from the semimetallic phase into spin-density-wave (SDW) and charge-density-wave (CDW) phases, respectively~\cite{herbut2009,araki2012,wu2013,janssen2014}.
More exotic interaction-induced states have also been discussed, such as Kekul\'e states~\cite{hou2007,roy2010,khari2012,classen2014}  or a topological Quantum (Spin) Hall state~\cite{raghu2008,grushin2013,daghofer2013,scherer2015}.
In fact, the nature of an ordered state crucially depends on the system's precise interaction profile, e.g., the magnitudes and ratios of the local and non-local short-ranged interaction terms.

Current experimental data suggests that free-standing graphene is in the semimetallic (SM)  phase~\cite{experiments,experiments2}.
From the theoretical side, calculations based on the constrained Random Phase Approximation (cRPA) and beyond provide values for the interaction parameters of the Coulomb repulsion for graphene and its few-layer relatives~\cite{wehling2011,rosner2015}.
Quantum Monte Carlo (QMC) studies for these parameters confirm the semimetallic behavior of physical graphene in agreement with the experimental findings~\cite{ulybyshev2013,smith2014}. 
At the same time, however, these results sugggest that the material may be not too far from a possible transition into an ordered state.
Other QMC calculations also find sizable charge-density and spin-current correlations, although they do not become long-ranged within the accessible parameter region~\cite{golor2015}.
Furthermore, a uniform and isotropic strain of about 15\% can be expected to induce an interaction-driven metal-insulator transition in graphene~\cite{assaad2015}.
It is therefore not inconceivable that physical graphene could possibly be tuned through a symmetry-breaking quantum phase transition~\cite{khvesh2001,herbut2006,Juricic2009,katanin2015}.
Similar conclusions may be expected to hold for other Dirac materials~\cite{wehling2014} and should also be relevant for ``artificial graphene''~\cite{polini2013}.

Despite the great progress in the last years, our theoretical understanding of the role of interactions in Dirac materials is far from being complete. 
In fact, QMC simulations typically suffer from a sign problem when nonlocal interaction parameters grow large, inducing a strong bias toward the antiferromagnetic state~\cite{ulybyshev2}.
%
%
Fermionic renormalization group approaches have provided important contributions to the understanding of interacting electrons in Dirac materials accounting for further-ranged interactions on equal footing~\cite{herbut2006,honerkamp2008,roy2009,classen2014,scherer2015}. 
These approaches are well-suited for the identification of the ordering tendencies and their classification by symmetries.
However, the purely fermionic approach typically misses important order-parameter fluctuations
and the description of symmetry-broken regimes in the phase diagram is intricate~\cite{metzner2012}.
An inclusion of order-parameter fluctuations aiming at more quantitative studies of the phase transitions and their critical behavior in Dirac materials can be achieved within bosonized approaches~\cite{rosa2001} that also allow to describe the symmetry-broken regime.
In this spirit, the SDW and CDW transitions have been investigated, however, only as completely separate transitions~\cite{herbut2009, janssen2014}.

Here, we take the vicinity of graphene-like materials to both the SDW and the CDW ordered states as a motivation to study the nature of the quantum multicritical point connected to the intersection of the different phase transition lines. 
%
\begin{figure}[t!]
\centering
 \includegraphics[width=0.8\columnwidth]{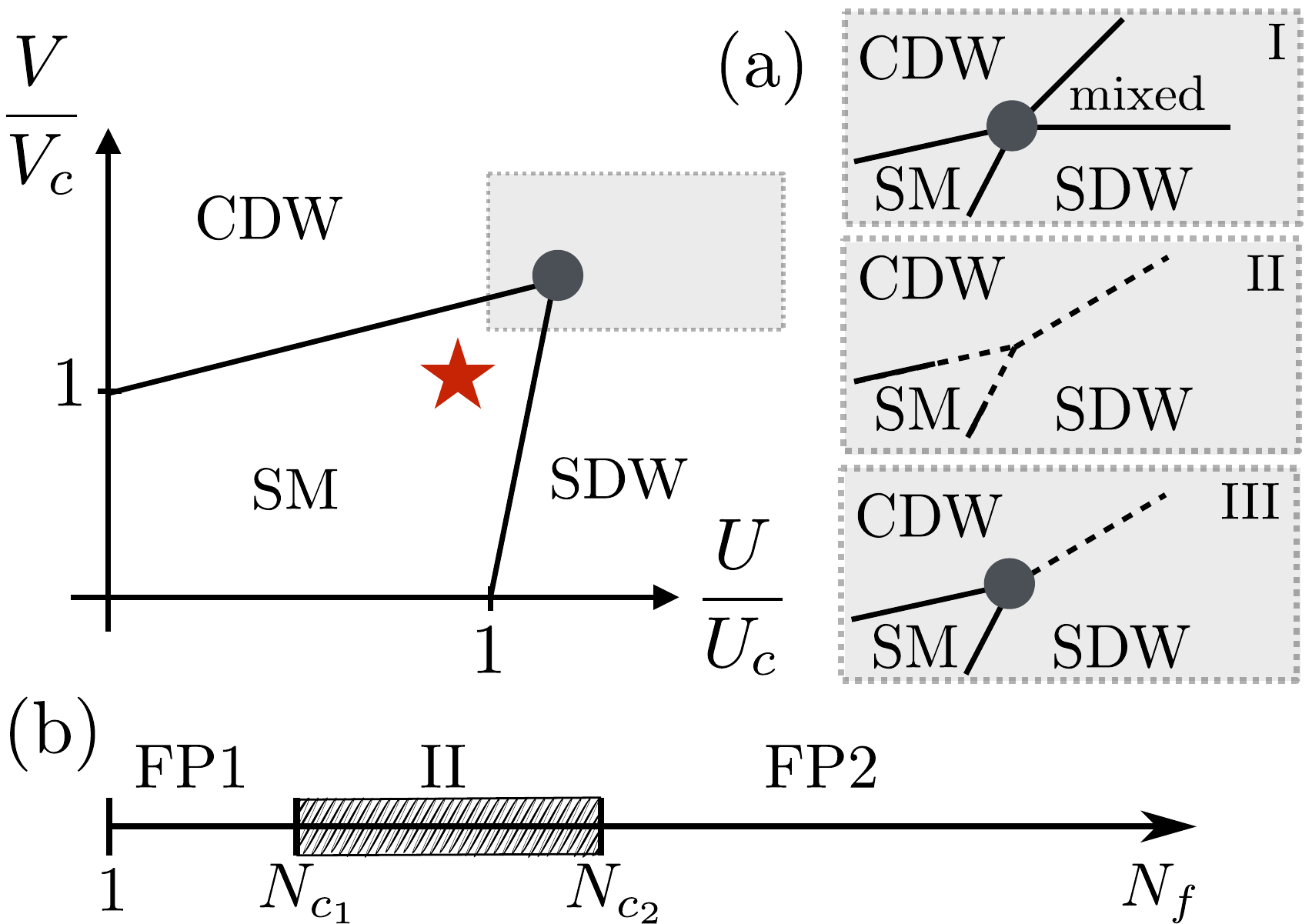}
 \caption{(a) Schematic phase diagram of the extended Hubbard model on the honeycomb lattice with onsite interaction $U$ and nearest-neighbor interaction $V$. Neutral suspended graphene is found to be in the semimetallic state indicated by the star. Solid lines denote second-order and dashed lines first-order transitions. The neighborhood of the multicritical point (gray shaded area) may be governed by either a (\RNum 1) second-order tetracritical point or a (\RNum 3) second-order bicritical point with first-order transition between the ordered states or by a (\RNum 2) first-order triple point.
 (b) Sketch of stability ranges of fixed points for generalized fermion flavor number $N_f$. Two different fixed points FP1 and FP2 are stable for small and large $N_f$, respectively. Graphene lies in the hatched region, where no stable fixed point exists. This leads to a first-order triple point in the phase diagram [situation (\RNum 2)]. The critical flavor numbers $N_{c_{i}}$ change considerably when including nonperturbative effects. In our approximation we find $N_{c_1}=1.6$ and $N_{c_2}=3.6$.}
\label{fig:phasediag}
\end{figure}
%
Such a study can be expected to reveal fascinating details of the phase diagram of Dirac materials, in particular, whether first-order or continuous phase transitions appear as a result of the competition of order parameters and whether there can be a coexistence of two ordered phases.
In principle, we can distinguish three different possibilities for the multicritical point: we can have either
(I) a second-order tetracritical point, allowing for the coexistence of the two ordered states, or (II) a triple point at which all transitions become first order, or (III) a second-order bicritical point with first-order transitions between the ordered states. A sketch of the phase diagram together with the three possible behaviors near the multicritical point is depicted in Fig.~\ref{fig:phasediag}(a).

For graphene and related materials multicritical behavior has previously been studied in different contexts using the $\epsilon$ expansion to first order, with $\epsilon$ being the distance from the upper critical space-time dimension of four~\cite{roy2011,roy2014,classen2015}.
Concerning the multicritical behavior and competition of CDW and SDW orders, we have recently put forward a corresponding study in Ref.~\cite{classen2015} using an effective Gross-Neveu-Yukawa model.
To first order in $\epsilon$, we have there found that a rather complex picture emerges as a function of the fermion ``flavor'' number $N_f$, i.e. the number of Dirac fermions.
The graphene case, $N_f=2$, appeared to be dominated by a second-order tetracritical point [situation (I) in Fig.~\ref{fig:phasediag}] with the universal behavior being in the same universality class as the SM-SDW transition, i.e., the ``chiral Heisenberg'' universality class~\cite{rosenstein1993}.
The first-order $\epsilon$ expansion is a formidable tool to detect and discover the qualitative aspects of these systems.
It may, however, be subject to considerable quantitative corrections when including higher orders~\cite{herbut1997,fei}. 
Furthermore, the convergence properties of the asymptotic series related to this type of expansion are {\it a priori} not clear, in particular when $\epsilon$ becomes of order one.

This paper aims at a considerable improvement in precision of our previous qualitative investigation by employing the functional renormalization group (FRG), which has proven to be a versatile and reliable tool to study both fermionic~\cite{rosa2001, janssen2012, mesterhazy2012, janssen2014, vacca2015} and bosonic systems~\cite{litim2011} at criticality.
In the context of multicritical behavior of bosonic systems with $\mathrm O(N_1) \oplus \mathrm O(N_2)$ symmetry, the significant quantitative improvement of the FRG approach as compared to the first-order $\epsilon$ expansion has been explicitly demonstrated~\cite{eichhorn2013}.

Employing the FRG, we are able to confirm the qualitative picture that we established in Ref.~\cite{classen2015}.
However, large quantitative modifications appear concerning the stability of the various fixed points as a function of $N_f$, with severe implications for the phase diagram in the graphene case, $N_f=2$.
We find that:
\begin{enumerate}[(1)]
 \item For small number of fermion flavors, $N_{f}<1.6$, the decoupled fixed point related to the antiferromagnetic transition (``chiral Heisenberg'' universality class) is stable.
 \item For intermediate fermion flavor numbers, $1.6< N_f < 3.6$, including the graphene case $N_f=2$,  there is no admissible stable fixed point, suggesting a triple point and corresponding first-order transitions.
 \item For large number of flavors, $N_f > 3.6$, we rediscover the novel stable fixed point with nontrivial interactions between the different sectors, found previously in the $\epsilon$ expansion~\cite{classen2015}.
\end{enumerate}
Our results concerning the ranges of stable fixed points are sketched in Fig.~\ref{fig:phasediag}(b). In the case where stable fixed points exist, we furthermore study the critical behavior in detail by investigating critical exponents and anomalous dimensions.
The rest of the paper is organized as follows: 
In the next section we introduce our effective model that couples the fermionic and bosonic degrees of freedom that become dominant in the vicinity of the multicritical point. 
We give an overview over the relevant terminology in Sec.~\ref{sec:critical} and explain our method in Sec.~\ref{sec:frg}.
In Sec.~\ref{sec:results} we discuss the resulting fixed-point structure and the concomitant critical behavior as function of space-time dimension and fermion flavor number $N_f$, and compare various limits to literature results.
In the limit of large $N_f$ we are able to present an analytic solution of the flow equations, including the full form of the fixed-point potential. 
The implications for the nature of the phase diagram are studied in Sec.~\ref{sec:phasediag} and we draw our conclusions in Sec.~\ref{sec:conclusion}.


\section{Extended Hubbard model and effective relativistic theory}

Let us start with the single-particle Hamiltonian for electrons with spin $s$ on the honeycomb lattice at half-filling,
\begin{align}
H_0&=-t\sum_{\mathbf{R},i,s}\left[u_s^\dagger(\mathbf{R}) v_s(\mathbf{R}+\boldsymbol{\delta}_i)+\text{h.c.}\right]\,,
\end{align}
summing over the sites $\mathbf{R}$ of the triangular sublattice and the three nearest-neighbor vectors $\boldsymbol{\delta}_i$. 
$u_s^{(\dagger)}$ and $v_s^{(\dagger)}$ correspond to annihilation (creation) operators on the two different sublattices of the honeycomb lattice.
This leads to two energy bands $\epsilon_{\mathbf{k}}=\pm t\left|\sum_{i=1}^3 \exp(i\, \mathbf{k}\cdot\boldsymbol{\delta}_i)\right|$ with linear and isotropic slope close to the pointlike Fermi surface located at the two inequivalent points $\mathbf{K}$, $\mathbf{K}'$ at the corners of the Brillouin zone.

Onsite and nearest-neighbor interactions are implemented by the interaction terms
\begin{align}\label{eq:int}
H_{I}=U\sum_{i}n_{i,\uparrow}n_{i,\downarrow}+V\sum_{\substack{\spitz{i,j},s,s'}}n_{i,s}n_{j,\pr{s}}\,,
\end{align}
with the density operators $n_{i,s}$ on site $i$.
Retaining the Fourier modes near $\mathbf{K}$, $\mathbf{K}'$ only, the low-energy model of the free electrons at temperature $T=0$ can be written as a relativistic Dirac field theory in the continuum~\cite{herbut2006}
\begin{align}
	S_F=\int d\tau dx^{D-1}\left[\bar\Psi\left(\mathbbm{1}_2\otimes\gamma_\mu\right)\partial_\mu\Psi\right]\,,
\end{align}
with space-time index $\mu=0,\dots,D-1$ and the $D$-dimensional derivative $\partial_\mu=(\partial_\tau, \nabla)$. The $(4\times4)$ gamma matrices obey the Euclidian Clifford algebra $\{\gamma_\mu,\gamma_\nu\}=2\delta_{\mu\nu}$. In $D=2+1$ dimensions they are explicitly represented by
$\gamma_0=\mathbbm{1}_2\otimes\sigma_z$, $\quad \gamma_1=\sigma_z\otimes\sigma_y$, $\quad \gamma_2=\mathbbm{1}_2\otimes\sigma_x$.
In this frame the spin-$\frac{1}{2}$ electrons and holes are described by an 8-component Dirac fermion $\Psi=\left(\Psi_\uparrow,\Psi_\downarrow\right)^T$ and its conjugate $\bar\Psi=\Psi^\dagger(\mathbbm{1}_2\otimes\gamma_0)$.
The Dirac field $\Psi$ is related to the Grassmann fields $u,v$ by 
\begin{align}
	 \Psi_s^\dagger(\mathbf x,\tau)&=\int\frac{d\omega d^{D-1}\mathbf q}{(2\pi)^{D}}e^{i\omega \tau+i \mathbf q\cdot \mathbf x}\big[
	u_s^\dagger(\mathbf K+\mathbf q,\omega),\nonumber\\
	& v_s^\dagger(\mathbf K+\mathbf q,\omega),u_s^\dagger(-\mathbf K+\mathbf q,\omega),v_s^\dagger(-\mathbf K+\mathbf q,\omega)
	\big]\label{eq:spinor}
\end{align}
with $\mathbf{K}'=-\mathbf{K}$. We can define two additional $(4\times4)$ matrices that anticommute with all $\gamma_\mu$: $\gamma_3=\sigma_x\otimes\sigma_y$ and $\gamma_5=\sigma_y\otimes\sigma_y$. Their product $\gamma_{35}=-i\gamma_3\gamma_5$ commutes with all $\gamma_\mu$, while it anticommutes with $\gamma_3$ and $\gamma_5$.

To describe the multicritical point in the phase diagram we introduce bosonic degrees of freedom related to the SDW and CDW fluctuations. These can be written in terms of the order-parameter fields~\cite{semenoff1984, herbut2006}
\begin{align} \label{eq:order-parameter}
	\Phi=\big(\chi,{\boldsymbol\phi}\big)=\left(\langle \bar\Psi\Psi\rangle,\langle \bar\Psi(\boldsymbol{\sigma}\otimes\mathbbm{1}_4)\Psi\rangle\right)\,,
\end{align}
which can also be understood as order parameters for the various possible chiral symmetries~\cite{gehring2015}.

Another very interesting set of order parameters, with possibly the \emph{same} quantum critical behavior~\cite{herbut2009}, is given by
\begin{equation}
 	\tilde\Phi=\big(\tilde\chi,\tilde{\boldsymbol\phi}\big)=\left(\langle \bar\Psi\gamma_{35}\Psi\rangle,\langle \bar\Psi(\boldsymbol{\sigma}\otimes\gamma_{35})\Psi\rangle\right)\,.
\end{equation}
These can be related to the much-discussed Quantum Anomalous and Quantum Spin Hall states~\cite{haldane1988}, and may also be relevant in the phase diagram of Dirac materials~\cite{raghu2008}.
Note that $\tilde{\boldsymbol\Phi}$'s zeroth component $\tilde \chi$ breaks the time-reversal symmetry, defined by the time-reversal operator~\cite{roy2009}
\begin{equation}
 T = (\sigma_y \otimes i \gamma_1 \gamma_5) K\,,
\end{equation}
where $K$ denotes complex conjugation, while its spatial part $\tilde{\boldsymbol\phi}$ breaks the SU(2) spin-rotational symmetry but respects time-reversal symmetry. In the following, we will focus on a condensation of the chiral order parameters $\chi$ and $\boldsymbol\phi$ only, cf. Eq.~\eqref{eq:order-parameter}, and assume that the fields $\tilde\chi$ and $\tilde{\boldsymbol\phi}$ are sufficiently massive so that their fluctuations become subdominant.

The spin-singlet Ising field $\chi\propto u_s^\dagger u_s - v_s^{\dagger} v_s$ characterizes the staggered density phase, i.e., the CDW state, whereas the Heisenberg triplet ${\boldsymbol\phi}\propto u_s^\dagger \boldsymbol{\sigma}_{ss'} u_{s'} - v_s^{\dagger} \boldsymbol{\sigma}_{ss'} v_{s'}$ corresponds to the antiferromagnetic SDW state. 
Near the putative transitions into the CDW and SDW states the fluctuations of the corresponding order parameters play a crucial role.
We incorporate their dynamics in the bosonic action
\begin{align}\label{eqn:actionB}
	\hspace{-0.2cm}S_B&\hspace{-0.1cm}=\hspace{-0.1cm}\int\hspace{-0.1cm} d\tau dx^{D-1}\Big[
	\frac{1}{2}\chi(-\partial_\mu^2+m_\chi^2)\chi+\frac{1}{2}{\boldsymbol\phi}(-\partial_\mu^2+m_\phi^2){\boldsymbol\phi}\nonumber\\
	&\quad\quad\quad+\frac{1}{8}\lambda_{2,0}\chi^4+\frac{1}{8}\lambda_{0,2}\big({\boldsymbol\phi}^2\big)^2+\frac{1}{4}\lambda_{1,1}\chi^2{\boldsymbol\phi}^2
	\Big],\,
\end{align}
where we also allow for a coupling $\lambda_{1,1}$ between the two order parameters. 
Finally, bosonic and fermionic degrees of freedom are coupled in terms of the Yukawa interactions
\begin{align}
	S_Y=\int d\tau dx^{D-1}\left[
	\bar{g}_{\chi}\chi\bar\Psi (\mathds{1}_2\otimes\mathds{1}_4)\Psi+\bar{g}_{\phi}{\boldsymbol\phi}\bar\Psi({\boldsymbol{\sigma}}\otimes\mathds{1}_4)\Psi\,
	\right].\nonumber
\end{align}
The complete action $S$ is then given by
\begin{align}\label{eq:mic}
	S=S_F+S_B+S_Y\,,
\end{align}
which respects Lorentz, spin-rotational, time-reversal and sublattice-exchange symmetry. The ordered phases are characterized by a finite expectation value of one or both bosonic fields, leading to the spontaneous breaking of the spin-rotational or sublattice-exchange symmetry.

In the following, it will prove useful to consider an arbitrary number $N_f$ of Dirac points in the spectrum, implemented by the replacement 
\begin{align}
\bar\Psi(\boldsymbol{\sigma}\otimes\mathbbm{1}_4)\Psi &\mapsto \bar\Psi(\boldsymbol{\sigma}\otimes\mathbbm{1}_{2N_f})\Psi\,, \\
\bar\Psi(\mathbbm{1}_2\otimes\mathbbm{1}_4)\Psi &\mapsto \bar\Psi(\mathbbm{1}_2\otimes\mathbbm{1}_{2N_f})\Psi\,,
\end{align}
where $\Psi$ and $\bar{\Psi}$ now have $2N_f$ components \emph{for each spin projection}. We will refer to $N_f$ as the fermion ``flavor'' number, with $N_f=2$ for graphene. Let us note that the explicit implementation of the flavor number is not important. To derive the results, only the Clifford algebra and the product $d_\gamma N_f$ is needed, where $d_\gamma$ denotes the dimension of the gamma matrices. We will also consider general space-time dimension $2<D<4$, with an eye on the physical $D=2+1$.
%


\section{Fixed points and critical behavior}\label{sec:critical}


\subsection{RG $\beta$ functions and fixed points}\label{sec:FPs}

Renormalization group theory describes the scale dependence of a physical system by providing $\beta$ functions for the different couplings of a theory.
The $\beta$ functions are differential equations encoding the evolution of the system with respect to the energy (or momentum) scale $k$.
Starting from a ``microscopic'' model for a system at some ultraviolet (UV) cutoff scale $k = \Lambda$, one can then infer the low-energy, or infrared (IR), characteristics in terms of the solution of the $\beta$ functions.
In our case, the UV scale $\Lambda$ corresponds to the scale at which our effective model, Eq.~\eqref{eq:mic}, is valid, and as such is much smaller than the bandwidth (at which an accurate lattice description would have to be employed).

More explicitly, we introduce the generalized set of dimensionless couplings for the theory by $\alpha_i$, $i \in \{1,2,\ldots \}$.
The $\beta$ functions can be written in the form $\partial_t \alpha_i=\beta_i(\alpha_1,\alpha_2,\ldots)$, where the change in scale is written in terms of the renormalization group time $t = \ln (k/\Lambda) \leq 0$. 
A fixed point~$\alpha^*$  of these equations is given by
\begin{align}
\beta_i(\alpha_1^\ast,\alpha_2^\ast,\ldots)=0 \quad \forall\ i,
\end{align}
and can be associated to a possible continuous phase transition.

The critical properties and scaling behavior near such a transition are encoded in the RG flow in the vicinity of the fixed point~$\alpha^*$,
\begin{align}
\partial_t \alpha_i=&B_{i,j} (\alpha_j^*-\alpha_j)+ \mathcal O\left((\alpha_j^*-\alpha_j)^2\right)\,,\end{align}
where $B_{i,j}=(\partial\beta_i/\partial \alpha_j)|_{\alpha=\alpha^*}$.
The eigenvalues $\theta_i$ of $(-B_{i,j})$ (``critical exponents'') are universal quantities that characterize the scaling laws at the putative continuous phase transition.
All positive critical exponents $\theta_i$ correspond to RG-relevant directions, i.e., the fixed point repels the flow in that direction. 
In turn, negative $\theta_i$ are RG irrelevant and correspond to attractive directions. 
Fixed points with no more than two relevant directions (corresponding to no more than two positive critical exponents, $\theta_1$ and $\theta_2$) can be accessed by tuning two microscopic parameters, e.g., onsite interaction $U$ and nearest-neighbor interaction $V$ for the microscopic theory, Eq.~\eqref{eq:int}, or the two masses $m_\chi^2$ and $m_\phi^2$ in our effective model, Eq.~\eqref{eq:mic}. In this work, we will call such fixed points ``stable''.
The third largest critical exponent $\theta_3$ then decides over the stability of a fixed point.
In addition, unitarity requires real Yukawa couplings $g_{\chi,\ast}, g_{\phi,\ast} \in \mathbbm{R}$, and the action has to be bounded from below, i.e., $\lambda_{2,0}^*, \lambda_{0,2}^*\geq 0$ and $\lambda^*_{1,1}>-\sqrt{\lambda^*_{2,0}\lambda_{0,2}^*}$.


\subsection{Classification of fixed points}\label{subsec:FPs}

As pointed out above, we are interested in the stable fixed point of the system, governing the quantum multicritical behavior of its phase diagram. 
The model incorporates the separate SM-to-SDW and the SM-to-CDW transitions described by the chiral Ising and chiral Heisenberg universality classes, respectively, as well as a purely bosonic model with a $\mathrm O(1)\oplus \mathrm O(3)$. 
Just as in such bosonic models, we can deduce the existence of some of the appearing fixed points and critical properties from symmetry considerations and from the limiting cases of the separate models~\cite{janssen2014,eichhorn2013}:
\begin{enumerate}[(1)]
\item The bosonic system, when the fermions completely decouple, i.e., $g_{\chi,\ast}^2=0$ and $g_{\phi,\ast}^2=0$, which puts the fermionic sector at its Gaussian fixed point. For the remaining bosonic sector there are three possible fixed points of the $\mathrm O(1)\oplus \mathrm O(3)$ model: the decoupled, the isotropic and the biconical one.
\item The chiral Ising sector with the Ising field $\chi$ at its nontrivial fermionic fixed point $g_{\chi,\ast}^2\neq 0$, and with the fermions decoupled from the Heisenberg field ${\boldsymbol\phi}$, $g_{\phi,\ast}^2 = 0$, which is then either at its bosonic Gaussian or Wilson-Fisher (Heisenberg) fixed point. The latter will be called ``chiral Ising plus Heisenberg'' (cI+H) fixed point in the following.
\item  The chiral Heisenberg sector with the Heisenberg field ${\boldsymbol\phi}$ at its nontrivial fermionic fixed point $g_{\phi,\ast}^2\neq 0$, and with the fermions decoupled from the Ising field $\chi$, $g_{\chi,\ast}^2 = 0$, which is then either at its bosonic Gaussian or Wilson-Fisher (Ising) fixed point. The latter will be called ``chiral Heisenberg plus Ising'' (cH+I) fixed point.
\end{enumerate}

Regarding the stability of these fixed points, we can infer the following: 
Every fixed point of the separate sectors will have one relevant direction related to its mass parameter.
The chiral Heisenberg and the chiral Ising fixed point do not show further relevant directions in the individual, uncoupled systems~\cite{janssen2014}.
In contrast, the Wilson-Fisher fixed point of the uncoupled sector $i$ ($i\in \{\chi,{\boldsymbol\phi}\}$), specified by ${g_i^{2}}^*=0$, features one additional relevant direction corresponding to the Yukawa coupling. 
But upon coupling this sector to the second bosonic field, this direction may or may not become irrelevant.
Thus, the cI+H and the cH+I are the most promising candidates for stable fixed points. 
Due to the Yukawa couplings being relevant below four space-time dimensions, the purely bosonic fixed points are unlikely to become stable by the coupling of both systems. 

This general expectation was indeed confirmed in the first-order $\epsilon$-expansion study of the coupled model, see Ref.~\cite{classen2015}, in which the cH+I fixed point appeared stable in the case of graphene ($N_f=2$).
On the other hand, for large $N_f$ a novel fixed point with both Yukawa interactions $g_{\chi,\ast}^2\neq 0$ and $g_{\phi,\ast}^2\neq 0$ became stable.
A third option, which was found in Ref.~\cite{classen2015} for intermediate $N_f$, is that there is no stable fixed point at all. 
In this case the flow does not exhibit scale-invariant behavior and the phase diagram close to the intersection of the various phase (SM, SDW and CDW) is goverened by a triple point with first-order transitions.


\section{Functional Renormalization}\label{sec:frg}


\subsection{Method}

We employ the functional renormalization group (FRG) to derive nonperturbative flow equations for the couplings of the quantum multicritical system~\cite{frg}.
The FRG provides a systematic approach to implement Wilson's idea of successively performing integration over degrees of freedom in the functional integral representation. 
It yields an exact functional differential equation describing the evolution of the generating functional for the one-particle irreducible correlation functions, i.e., the effective action $\Gamma$, with an infrared momentum scale $k$, reading~\cite{wetterich1993}
\begin{align}\label{eqn:Wetterich}
\partial_t\Gamma_k = \frac{1}{2}\text{STr}\eck{(\Gamma_k^{(2)}+R_k)^{-1}\partial_t R_k}.
\end{align}
The scale-$k$-dependent or \emph{flowing} action $\Gamma_k$ interpolates between the microscopic action $\Gamma_{k\rightarrow\Lambda}=S$ at UV cutoff $\Lambda$ and the full quantum effective action  $\Gamma_{k\rightarrow0}$ in the IR. 
To ensure the interpolation, we have introduced the regulator function $R_k$.
It induces the iterative integration procedure and ensures that only modes with high momentum $\abs{q}\gtrsim k$ give a contribution to the integral in $\Gamma_k$, thereby avoiding infrared singularities. 
Therefore it has to satisfy the requirements $R_k(q)\rightarrow \infty$ for $k\rightarrow \Lambda \rightarrow \infty$ and $R_k(q)\rightarrow 0$ for $k/|q|\rightarrow 0$.
%
%

Explicitly, the regulator modifies the microscopic action which appears in the functional integral representation of the partition function, $Z=\int_\Lambda \mathcal{D}\varphi\, e^{-S[\varphi]}$, by replacing
\begin{align}
S\rightarrow S &+ \int\frac{d^Dqd^Dp}{(2\pi)^{2D}}\left[\frac{1}{2}\chi(-q)R_{\chi,k}^{(B)}(q,p)\chi(p)\right. \nonumber\\
&\hspace{-0.2cm}\left.+\frac{1}{2}{\boldsymbol\phi}(-q)R_{\phi,k}^{(B)}(q,p){\boldsymbol\phi}(p) + \bar\Psi(q)R_k^{(F)}(q,p)\Psi(p)\right].\nonumber
\end{align}
The flowing action is then defined as the Legendre transform of the regularized Schwinger functional $W_k[J]=\ln Z_k$, see, for instance, Ref.~\cite{frg} for details.

Further, in Eq.~(\ref{eqn:Wetterich}) (the so-called ``Wetterich'' equation) we have abbreviated $\partial_t=k\partial_k$, using the renormalization group time $t = \ln(k/\Lambda)$. The Hessian $\Gamma_k^{(2)}$ is given by
\begin{align}
\rund{\Gamma_k^{(2)}}_{i,j}(p,q)=\frac{\overrightarrow{\delta}}{\delta\Phi(-p)^T}\Gamma_k\frac{\overleftarrow{\delta}}{\delta\Phi(q)}\,,
\end{align}
with $\Phi=(\chi,{\boldsymbol\phi},\Psi,\bar\Psi^T)^T$ representing a collective field variable for all fermionic and bosonic degrees of freedom of our model~\eqref{eq:mic}.

In the Wetterich equation, the regulators for this model are combined to
\begin{align}
R_k=\begin{pmatrix}R_{\chi,k}^{(B)} & 0 & 0 & 0\\ 0  & R_{\phi,k}^{(B)} & 0 & 0 \\ 0 & 0 & 0 & R_k^{(F)} \\ 0 & 0 & -R_k^{(F)T} & 0 \end{pmatrix}\,.
\end{align}
and STr sums over all degrees of freedom including a minus sign in the fermionic sector as well as a loop integration over momenta.

The FRG method provides a unified framework to access universal as well as nonuniversal properties of physical systems. 
It may be employed to describe the critical behavior in the vicinity of continuous classical or quantum phase transitions and is also applicable to systems away from criticality.
By means of suitable expansion schemes it has also been used to study first-order phase transitions~\cite{frg}.
The FRG can be applied in arbitrary (fractional) dimension, and even low-order truncations already appear to give reasonably accurate results in both purely bosonic as well as coupled boson-fermion systems~\cite{litim2011, rosa2001, janssen2014}.
%


\subsection{Truncation}

While the Wetterich equation itself is an exact identity, it can usually not be solved exactly. In this work, we use a scheme inspired by the derivative expansion, which we truncate after the leading order. Explicitly, we employ the following ansatz for $\Gamma_k$ [so-called ``improved local potential approximation'' (LPA')],
\begin{align} \label{eq:truncation}
	\Gamma_k=&\int d^Dx~\Big(
	Z_{\Psi,k}\bar\Psi(\mathds{1}_2\otimes \gamma_\mu)\partial_\mu\Psi\nonumber\\
	&-\frac{1}{2}Z_{\chi,k}\chi\partial_\mu^2\chi-\frac{1}{2}Z_{\phi,k}{\boldsymbol\phi}\partial_\mu^2{{\boldsymbol\phi}}+ U_k(\bar\rho_\chi,\bar\rho_\phi) \nonumber\\
	&+ \bar{g}_{\chi, k}\chi\bar\Psi (\mathds{1}_2\otimes\mathds{1}_4)\Psi+\bar{g}_{\phi, k}{\boldsymbol\phi}\bar\Psi ({\boldsymbol{\sigma}}\otimes\mathds{1}_4)\Psi \Big)\,,
\end{align}
which is a direct generalization of the quantitatively successful truncation used for the separate chiral Heisenberg and chiral Ising universality classes~\cite{rosa2001, janssen2014}.
In the first line of Eq.~\eqref{eq:truncation}, we have introduced the kinetic part of the fermion fields, followed by the kinetic parts of the order-parameter fields in the second line. We assume scale-dependent, but field-independent wave-function renormalizations $Z_{\Psi,k}$, $Z_{\chi,k}$ and $Z_{\phi,k}$. 
In the third line, the Yukawa couplings also become scale-dependent quantities.
The scale-dependent effective bosonic potential $U_k$, also appearing in the second line, only depends on the invariants $\rho_\varphi=\frac{1}{2}\varphi^2$, $\varphi \in \{\chi,{\boldsymbol\phi}\}$, as imposed by the symmetry of the original action, Eq.~\eqref{eq:mic}.
For most purposes, we will in the following expand the effective potential about its scale-dependent minimum $(\bar\rho_{\chi,\text{min}}$, $\bar\rho_{\phi,\text{min}})$, the latter being either zero or positive, then describing a regime of spontaneously broken symmetry.
In case a nonvanishing $(\bar\rho_{i,\text{min}}$ survives the integration towards the IR, $k\rightarrow 0$, it corresponds to a finite vacuum expectation value for the order-parameter fields $\chi$ and/or ${\boldsymbol\phi}$, i.e., an ordered phase.
%


\subsection{Flow equations}
\label{sec:frgflow}


\subsubsection{Effective potential}

In order to determine the scaling behavior of the effective action, we will consider dimensionless quantities in the following. Therefore, we define the dimensionless version of the effective potential
\begin{align}
u(\rho_\chi,\rho_\phi)=k^{-d}U\left(\frac{k^{D-2}}{Z_{\chi,k}}\bar\rho_\chi,\frac{k^{D-2}}{Z_{\phi,k}}\bar\rho_\phi\right)\,,
\end{align}
and the corresponding Yukawa couplings
\begin{align}
g_{\chi/\phi}^2=\frac{k^{D-4}}{Z_{\chi/\phi,k}Z_{\Psi,k}^2}\bar g_{\chi/\phi,k}^2\,.
\end{align}
We also define the anomalous dimensions
\begin{align}\label{eqn:anom}
\eta_{\chi/\phi}=-\frac{\partial_t Z_{\chi/\phi,k}}{Z_{\chi/\phi,k}} \quad\text{and}\quad \eta_{\Psi}=-\frac{\partial_t Z_{\Psi,k}}{Z_{\Psi,k}}.
\end{align}

To obtain the flow equation for the dimensionless scale-dependent effective potential $u$, Eq.~(\ref{eqn:Wetterich}) is evaluated for constant bosonic fields $\chi$, ${\boldsymbol\phi}$ and vanishing fermion fields $\Psi$. The resulting flow equation can be compactly written as
\begin{align}\label{eq:potflow}
	\partial_t u=&(D-2+\eta_\chi)\rho_\chi u^{(1,0)}+(D-2+\eta_\phi)\rho_\phi u^{(0,1)}-D u\nonumber\\
	&+I_{R}(\omega_\chi,\omega_\phi,\omega_{\phi\chi}^2)+2I_{G}(u^{(0,1)})\nonumber\\
	&-2N_f\Big[I_\psi(\omega_\psi^+)+I_\psi(\omega_\psi^-)\Big]\,,
\end{align}
where we have defined the following quantities
\begin{eqnarray}
	u^{(i,j)}&=&\frac{\partial^{i+j} }{\partial \rho_\chi^i \partial \rho_\phi^j}u(\rho_\chi,\rho_\phi)\,,\\
	\omega_\chi&=&u^{(1,0)}+2\rho_\chi u^{(2,0)}\,,\\
	\omega_\phi&=&u^{(0,1)}+2\rho_\phi u^{(0,2)}\,,\\
	\omega_{\phi\chi}^2&=&4\rho_\phi \rho_\chi \big(u^{(1,1)}\big)^2\,,\\
	\omega_\psi^{\pm}&=&2\rho_\chi g_\chi^2+2\rho_\phi g_\phi^2\pm 4\sqrt{\rho_\chi \rho_\phi}g_\chi g_\phi\,.
\end{eqnarray}
The {\it threshold} functions $I_R, I_G$, and $I_\Psi$ involve the loop integrations and the regulator dependence.
For a suitable choice of the regulator functions for the bosons and fermions, these integrations can be performed analytically and the result can be given in a closed form, see Appendix~\ref{app:thresh}.

The effective dimensionless potential $u$ is expanded about its scale-dependent dimensionless minimum at $(\kappa_\chi,\kappa_\phi)=((k^{D-2}/Z_\chi )\bar\rho_{\chi,\text{min}}, (k^{D-2}/Z_\phi) \bar\rho_{\phi,\text{min}})$. Its IR limit  corresponds to the vacuum expectation values of $\chi$ and ${\boldsymbol\phi}$, $\lim_{k\rightarrow0}\kappa_{\varphi}=\langle\frac{1}{2}\varphi^2\rangle$, $\varphi\in\{\chi,\phi\}$.
We may distinguish four qualitatively different combinations for the location of the minimum of $u$: 
\begin{enumerate}[(i)]
\item Both sectors remain in the symmetric regime (SYM-SYM) with $(\kappa_\chi,\kappa_\phi)=(0,0)$, or 
\item either of the symmetries is spontaneously broken $\kappa_\chi\neq0, \kappa_\phi=0$~(SSB-SYM) or vice versa (SYM-SSB), or 
\item both order parameters attain a nonzero vacuum expectation value $\kappa_{\chi,\phi}\neq0$ (SSB-SBB). 
\end{enumerate}
The following parameterization of the effective potential in terms of a two-dimensional Taylor expansion accounts for all of these scenarios
\begin{align}
u(\rho_\chi,\rho_\phi)= \sum_{m+n\geq1}^{m+n=N}\frac{\lambda_{n,m}}{n!m!}(\rho_\chi-\kappa_\chi)^n(\rho_\phi-\kappa_\phi)^m\,,
\end{align}
with $\lambda_{1,0}=m_\chi^2$ if $\kappa_\chi=0$ and $\lambda_{1,0}=0$ if $\kappa_\chi\neq0$. 
Analogous definitions are used for $\lambda_{0,1}, \kappa_\phi$ and $m_\phi^2$. 
The $\beta$ functions for the expansion parameters in the different regimes are then obtained by the projections:
\begin{enumerate}[(i)]
\item SYM-SYM regime:
\begin{align}
\partial_t\lambda_{n,m}=&(\partial_tu)^{(n,m)}|_{\begin{subarray}{l}\rho_\chi=0\\\rho_\phi=0\end{subarray}}\,.
\end{align}
\item SSB-SYM regime:
\begin{align}
\label{eqn:SSBSYM1}
\partial_tm_\phi^2&=\left.(\partial_tu)^{(0,1)}+\lambda_{1,1}\partial_t\kappa_\chi\right|_{\begin{subarray}{l}\rho_\chi=\kappa_\chi\\\rho_\phi=0\end{subarray}}\,,\\
\partial_t\kappa_\chi&=-\left.\frac{(\partial_tu)^{(1,0)}}{\lambda_{2,0}}\right|_{\begin{subarray}{l}\rho_\chi=\kappa_\chi\\\rho_\phi=0\end{subarray}}\,,\\
\partial_t\lambda_{n,m}&=\left.(\partial_tu)^{(n,m)}+\lambda_{n+1,m}\partial_t\kappa_\chi\right|_{\begin{subarray}{l}\rho_\chi=\kappa_\chi\\\rho_\phi=0\end{subarray}}\,.\label{eqn:SSBSYM2}
\end{align}
The projections of the SYM-SSB regime can be obtained accordingly by exchanging $\chi$ and ${\boldsymbol\phi}$ in Eq.~(\ref{eqn:SSBSYM1}) - (\ref{eqn:SSBSYM2}). 
\item SSB-SSB regime:
\begin{align}
\partial_t\kappa_\chi&=\left.\frac{\lambda_{1,1}(\partial_tu)^{(0,1)}-\lambda_{0,2}(\partial_tu)^{(1,0)}}{\lambda_{2,0}\lambda_{0,2}-\lambda_{1,1}^2}\right|_{\begin{subarray}{l} \rho_\chi=\kappa_\chi\\ \rho_\phi=\kappa_\phi\end{subarray}}\,,\\
\partial_t\kappa_\phi&=\left.\frac{\lambda_{1,1}(\partial_tu)^{(1,0)}-\lambda_{2,0}(\partial_tu)^{(0,1)}}{\lambda_{2,0}\lambda_{0,2}-\lambda_{1,1}^2}\right|_{\begin{subarray}{l}\rho_\chi=\kappa_\chi\\\rho_\phi=\kappa_\phi\end{subarray}}\,,\\
\partial_t\lambda_{n,m}&=\Big[(\partial_tu)^{(n,m)} + \lambda_{n+1,m}\partial_t\kappa_\chi \nonumber\\
&\quad\quad+ \left.\lambda_{n,m+1}\partial_t\kappa_\phi\Big]\right|_{\begin{subarray}{l}\rho_\chi=\kappa_\chi\\\rho_\phi=\kappa_\phi\end{subarray}}\,.
\end{align}
\end{enumerate}

For our numerical results we expand the effective potential up to order $\chi^8,{\boldsymbol\phi}^8$ (LPA' 8) and check the convergence of critical quantities with respect to the inclusion of higher orders in the fields up to order $\chi^{12},{\boldsymbol\phi}^{12}$ (LPA' 12).
An interesting option to overcome the limitations of local expansions in field space makes use of pseudo-spectral methods. This has recently been put forward within the FRG approach to access fixed points and critical exponents, see Ref.~\cite{borchardt2015}.
We leave the implementation of these methods in the present context for future work.


\subsubsection{Yukawa couplings}

For the projection of the flow of the Yukawa couplings we split the two-point function into its fluctuation dependent and independent parts
$
\Gamma_{k,0}^{(2)}=\left.\Gamma_k^{(2)}\right|_{\chi={\boldsymbol\phi}=\Psi=0},
$ and
$
\Delta\Gamma_k^{(2)}=\Gamma_k^{(2)}-\Gamma_{k,0}^{(2)}\,.
$
Then we expand the Wetterich equation in the following way
\begin{align}\label{eqn:logWetterich}
\partial_t\Gamma_k =& \frac{1}{2}\tilde\partial_t\text{STr}\eck{\ln(\Gamma_k^{(2)}+R_k)} \\
=&\frac{1}{2}\tilde\partial_t\text{STr}\eck{\ln(\Gamma_{k,0}^{(2)}+R_k)}  \nonumber\\
&+ \frac{1}{2}\tilde\partial_t\text{STr}\sum_{n=1}^\infty\frac{(-1)^{n+1}}{n}\eck{(\Gamma_{k,0}^{(2)}+R_k)^{-1}\Delta\Gamma_k^{(2)}}^n\nonumber\,.
\end{align}
Here, the scale derivative $\tilde\partial_t$ acts only on the $t$-dependence of the regulator.
The fields are divided into their vacuum expectation value and a fluctuating part, $\chi=\chi_0+\Delta\chi$, ${\phi}_3=\phi_{3,0}+\Delta\phi_3$, and $\phi_{1,2}=\Delta\phi_{1,2}$. This allows us to devise suitable projection rules to extract the flow of Yukawa couplings
\begin{widetext}
\begin{align}
\partial_t g_\chi&=\frac{1}{6 N_f d_\gamma}\text{Tr}\eck{\frac{\overrightarrow\delta}{\delta \Delta\chi(p')}\frac{\overrightarrow\delta}{\delta\bar\Psi(p)}\tilde\partial_t\text{STr}\eck{\rund{\frac{\Delta\Gamma_k^{(2)}}{\Gamma_{k,0}^{(2)}+R_k}}^3}\frac{\overleftarrow\delta}{\delta\Psi(q)}}_{\begin{subarray}{l}p'=p=q=0\\\bar\Psi=\Psi=\Delta\chi=\Delta\phi=0\end{subarray}}\,,\\
\partial_t g_\phi&=\frac{1}{6 N_f d_\gamma}\text{Tr}\eck{(\sigma_1\otimes\mathds{1}_4)\frac{\overrightarrow\delta}{\delta \phi_1(p')}\frac{\overrightarrow\delta}{\delta\bar\Psi(p)}\tilde\partial_t\text{STr}\eck{\rund{\frac{\Delta\Gamma_k^{(2)}}{\Gamma_{k,0}^{(2)}+R_k}}^3}\frac{\overleftarrow\delta}{\delta\Psi(q)}}_{\begin{subarray}{l}p'=p=q=0\\\bar\Psi=\Psi=\Delta\chi=\Delta\phi=0\end{subarray}}\,.
\end{align}
\end{widetext}
The resulting $\beta$-functions can again be calculated analytically and presented in a closed form. 
For these Yukawa couplings, the expressions, however, are rather lengthy and are therefore deferred to Appendix~\ref{app:betafcts}. 


\subsubsection{Anomalous dimensions}

We finally need a projection prescription for $\partial_t Z_i$, $i\in\{\chi,{\boldsymbol\phi},\psi\}$. 
To this end, the expansion of the Wetterich equation, Eq.~\eqref{eqn:logWetterich}, is evaluated for momentum-dependent fields to calculate the anomalous dimensions according to Eq.~(\ref{eqn:anom}).
Here, we choose to project onto the Goldstone modes.
Note that a projection onto the radial mode would provide admixtures of additional terms.
Details on this choice are presented in Appendix~\ref{app:projection}.
For the anomalous dimension of the Heisenberg field we can arbitrarily select one of the two Goldstone modes in order to determine $\partial_t Z_\phi$. 
For the Ising field, however, there is no Goldstone.
Here, we define $\eta_\chi$ as the limiting case $N \to 1$ of $N$ copies of the Ising field, see Appendix~\ref{app:projection} for details. 
In summary, we use the following prescriptions:

\begin{widetext}
\begin{align}
\partial_t Z_\psi&=\left.\frac{i}{4DN_fd_\gamma}\text{Tr}\eck{(\mathds{1}_2\otimes\gamma_\mu)\frac{\partial}{\partial p_\mu}\int\frac{d^Dq}{(2\pi)^D}\frac{\overrightarrow\delta}{\delta\bar\Psi(p)}\tilde\partial_t\text{STr}\eck{\rund{\frac{\Delta\Gamma_k^{(2)}}{\Gamma_{k,0}^{(2)}+R_k}}^2}\frac{\overleftarrow\delta}{\delta\bar\Psi(q)}}\right|_{\begin{subarray}{l}p=q=0 \\ \Delta\chi=\Delta\phi=\bar\Psi=\Psi=0\end{subarray}}\,.
\end{align}
and
\begin{align}
\partial_t Z_\chi&=\lim_{N\rightarrow1}\left[\left.-\frac{1}{4}\frac{\partial}{\partial p^2}\int\frac{d^Dq}{(2\pi)^D}\frac{\overrightarrow\delta}{\delta\Delta\chi_{G,1}(-p)}\tilde\partial_t\text{STr}\eck{\rund{\frac{\Delta\Gamma_k^{(2)}}{\Gamma_{k,0}^{(2)}+R_k}}^2}\frac{\overleftarrow\delta}{\delta\Delta\chi_{G,1}(q)}\right|_{\begin{subarray}{l}p=q=0 \\ \Delta\chi=\Delta\phi=\bar\Psi=\Psi=0\end{subarray}} \right] \label{eqn:etachi}\,,\\
\partial_t Z_\phi&=\left.-\frac{1}{4}\frac{\partial}{\partial p^2}\int\frac{d^Dq}{(2\pi)^D}\frac{\overrightarrow\delta}{\delta\Delta\phi_1(-p)}\tilde\partial_t\text{STr}\eck{\rund{\frac{\Delta\Gamma_k^{(2)}}{\Gamma_{k,0}^{(2)}+R_k}}^2}\frac{\overleftarrow\delta}{\delta\Delta\phi_1(q)}\right|_{\begin{subarray}{l}p=q=0 \\ \Delta\chi=\Delta\phi=\bar\Psi=\Psi=0\end{subarray}}\,,
\end{align}
\end{widetext}
with $\chi\rightarrow(\chi_R+\Delta\chi_{R,1},\Delta\chi_{G,1},\ldots,\Delta\chi_{G,N-1})$.
The full analytical expressions for the anomalous dimensions in terms of threshold functions are given in Appendix~\ref{app:anom}. 
Further, in Appendix~\ref{app:threshold}, the corresponding threshold functions are listed for general regulator, as well as explicit expressions for the linear and sharp regulators are given. 

In this study, we use the linear regulator to calculate critical exponents and the sharp regulator to determine the perturbative limit of the above flow equations. 
We use this as a crosscheck in the following way: 
The upper critical space-time dimension of the theory is four. 
Consequently, in $D=4-\epsilon$ dimensions, perturbation theory becomes reliable. 
The one-loop flow equations obtained in a standard Wilsonian approach can be reproduced from the FRG approach as a limiting case. 
To this end, we consider the symmetric regime and neglect all perturbatively irrelevant operators in the ansatz for the effective action.
Then, expanding the flow equations in $\epsilon=4-D$ yields exactly the 1-loop results of Ref.~\cite{classen2015}.


\section{Results}\label{sec:results}

Using the FRG flow equations, we can now search for fixed points and study their evolution as a function of space-time dimension $D$ and number of fermion flavors $N_f$.
%
%
We start with benchmarking by comparing to the results on the separate chiral Ising and the chiral Heisenberg universality classes from Ref.~\cite{janssen2014,classen2015}.%
\footnote{We have also checked the existence and properties of the purely bosonic fixed points, which can be compared with Ref.~\cite{eichhorn2013}---however, since these fixed points turn out to have more than three relevant directions when fermions are present (as expected), they will not play a role in the remainder of this study.}


\subsection{Chiral Ising and chiral Heisenberg universality class for $N_f=2$}

\begin{table}[b]
\caption{\label{Tab:01} Anomalous dimensions and largest critical exponents from this work (first line) in comparison with different methods and models for the chiral Ising universality class, $N_f=2$ and $D=3$. The boldface numbers show the values of the decisive third critical exponent, exhibiting that this multicritical fixed point is unstable.}
\label{tab:I+H}
\begin{tabular*}{\linewidth}{@{\extracolsep{\fill} } c c  c c c   c c c}
\hline\hline
model & method& $\theta_1$ & $\theta_2$&$\theta_3$  & $\eta_\chi$ & $\eta_\phi$ & $\eta_\psi$\\
\hline
cI+H & FRG &1.359 & 0.983 & {\bf 0.719} & 0.760 &  0.041 &  0.032 \\
&$\epsilon^1$ exp~\cite{classen2015} &1.545 & 1.048 & {\bf 0.571} & 0.571 & 0 &  0.071\\ 
\hline
cI & FRG~\cite{janssen2014} &  & 0.982 &  & 0.760 &   &  0.032 \\
& FRG~\cite{vacca2015} & & 0.996 & & 0.789 & & 0.031 \\
& $\epsilon^2$ exp~\cite{rosenstein1993} &  & 1.055 &  & 0.695 &   &  0.065 \\
& MC~\cite{karkkainen1994} &  & 1.00 &  & 0.754 &   &   \\
\hline
O(3)& $\epsilon^5$ exp~\cite{guida1998} &1.419 &  &  &  &  0.037 & \\
& MC~\cite{campostrini2002} &1.406 &  &  &  &  0.038 & \\
\hline\hline
\end{tabular*}
\end{table}
%
\begin{table}[t]
\caption{\label{tab:H+I} Anomalous dimensions and largest critical exponents  from this work (first line) in comparison with different methods and models for the chiral Heisenberg universality class, $N_f=2$ and $D=3$. The boldface numbers show the values of the decisive third critical exponent. Note the dramatic change in $\theta_3$ for the cH+I when going from the $\epsilon$ expansion to the FRG results, rendering the cH+I fixed point unstable.}
\begin{tabular*}{\linewidth}{@{\extracolsep{\fill} } c  c  c c c   c c c}
\hline\hline
model & method & $\theta_1$ & $\theta_2$&$\theta_3$  & $\eta_\chi$ & $\eta_\phi$ & $\eta_\psi$\\
\hline
cH+I & FRG &1.564 & 0.773 & {\bf 0.241} & 0.044 &  1.015 &  0.084 \\
& $\epsilon^1$ exp \cite{classen2015} &1.667 & 0.473 & {\bf -0.8} & 0 & 0.8 &  0.3 \\
\hline
cH & FRG \cite{janssen2014} &  & 0.772 &  &  &  1.015 &  0.084 \\
& $\epsilon^2$ exp \cite{rosenstein1993} &  & 0.834 &  &  &  0.959 &  0.242 \\
& MC \cite{toldin2015} &  & 1.19 &  &  &  0.70 &   \\
\hline
O(1)& $\epsilon^5$ exp \cite{guida1998} &1.590 &  &  & 0.036 &   & \\
& MC \cite{hasenbusch2011} &1.587 &  &  & 0.036 &   & \\
\hline\hline
\end{tabular*}
\end{table}
%

%
In Tables \ref{tab:I+H} and \ref{tab:H+I}, we give our best estimates for the critical exponents of the chiral Ising plus Heisenberg (cI+H) and the chiral Heisenberg plus Ising (cH+I) fixed points, respectively, for $N_f=2$ and $D=3$.
As an important result, we find that both of these fixed points exhibit three relevant directions and are therefore unstable. 
This finding is different from the leading-order $\epsilon$ expansion, in which the cH+I fixed point appeared stable, featuring only two relevant directions. 
Since these fixed points are composites from a chiral and a purely bosonic model, we can compare several quantities with previous (FRG) calculations and other methods. 
The exponent $\theta_1$ is given by the correlation length exponent in a O(1) or a O(3) model for the cH+I and the cI+H fixed point, respectively. 
The second critical exponent $\theta_2$ is inherited from the chiral Heisenberg (chiral Ising) model. 
Additionally, the values of the anomalous dimensions are inherited from the separate models, cf.~Table~\ref{tab:I+H}.
In the case of the cI+H fixed point, the anomalous dimensions $\eta_\chi$ and $\eta_\psi$ come from the chiral Ising model, and $\eta_\phi$ from the bosonic O(3) (Heisenberg) model. 
For the cH+I fixed point, $\eta_\phi$ and $\eta_\psi$ can be inferred from the chiral Heisenberg system, while $\eta_\chi$ is adopted from the bosonic O(1) (Ising) model.
As can be read off from Table~\ref{tab:I+H} (Table~\ref{tab:H+I}), we find good agreement between our estimates with the well-known results for the bosonic Heisenberg (bosonic Ising) universality class, i.e., for the exponents $\theta_1$ and $\eta_\phi$ ($\eta_\chi$).
For the chiral Ising (chiral Heisenberg) universality class, the exponents [$\theta_2$ and $\eta_\chi$ ($\eta_\phi$)] are not precisely known. In any case, we find very good agreement with previous FRG computations~\cite{janssen2014, vacca2015}, and reasonably good agreement with the second-order $\epsilon$-expansion results~\cite{rosenstein1993}. When comparing our estimates with the Monte-Carlo results, we find good agreement in the case of the chiral Ising universality class~\cite{karkkainen1994}, whereas in the case of the chiral Heisenberg universality class we observe significant discrepancies to the newest Monte-Carlo estimates~\cite{sorella2012, assaad2013, toldin2015}, see Ref.~\cite{janssen2014} for a discussion.

Let us note that the exponent $\theta_3$ for the decoupled fixed point in the purely bosonic $\mathrm O(N_1)\oplus \mathrm O(N_2)$ model can be deduced from $\theta_1$ and $\theta_2$ by an exact scaling relation~\cite{calabrese2003}.
An equivalent relation is at present unknown for the fermionic model studied here, because there is no {\it a priori} knowledge about the scaling dimensions of the fermionic operators.


\subsection{From $\bf{4-\epsilon}$ to 3 space-time dimensions}

\begin{figure}[t!]
\centering
 \includegraphics[width=.9\columnwidth]{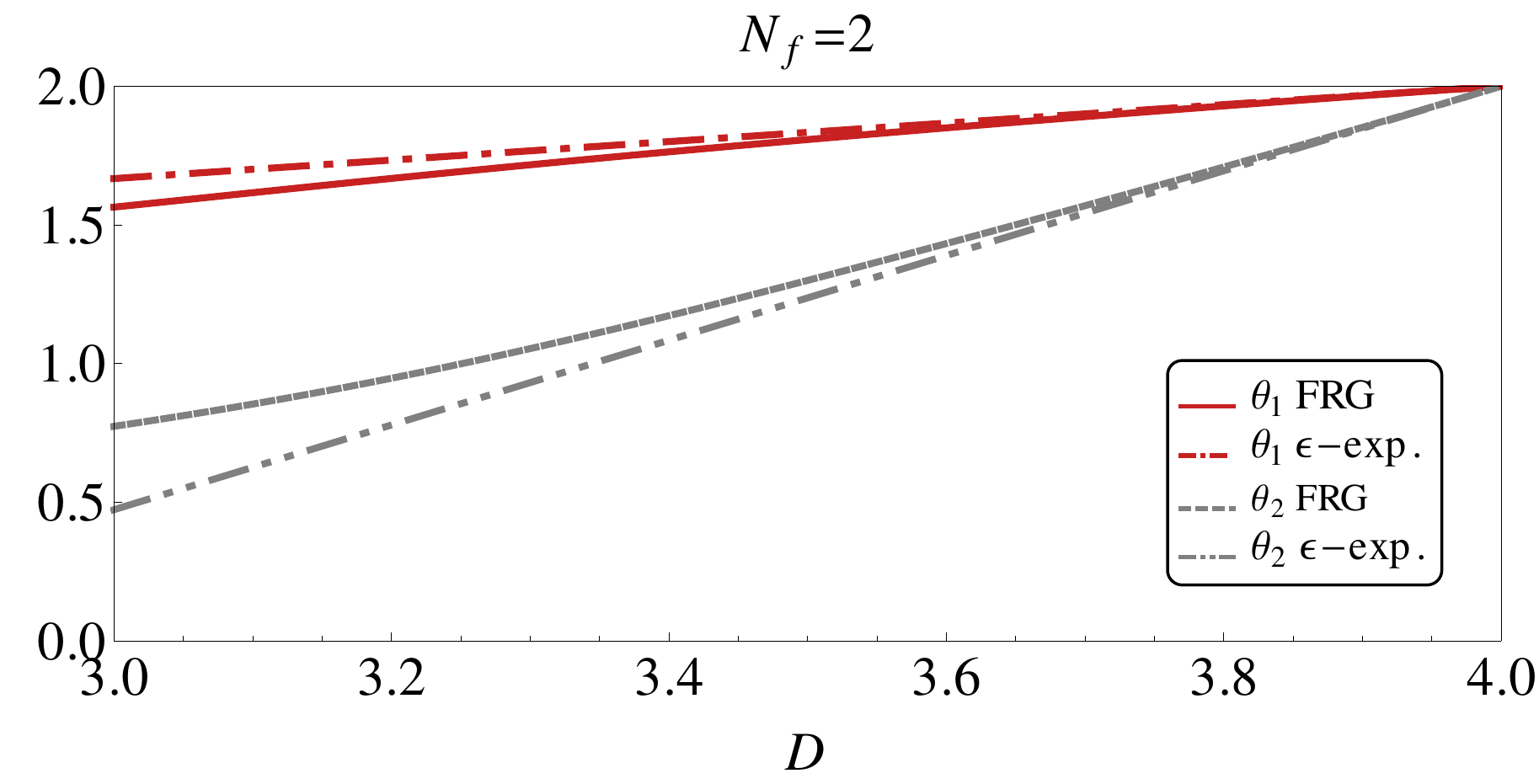}\\
 \includegraphics[width=.9\columnwidth]{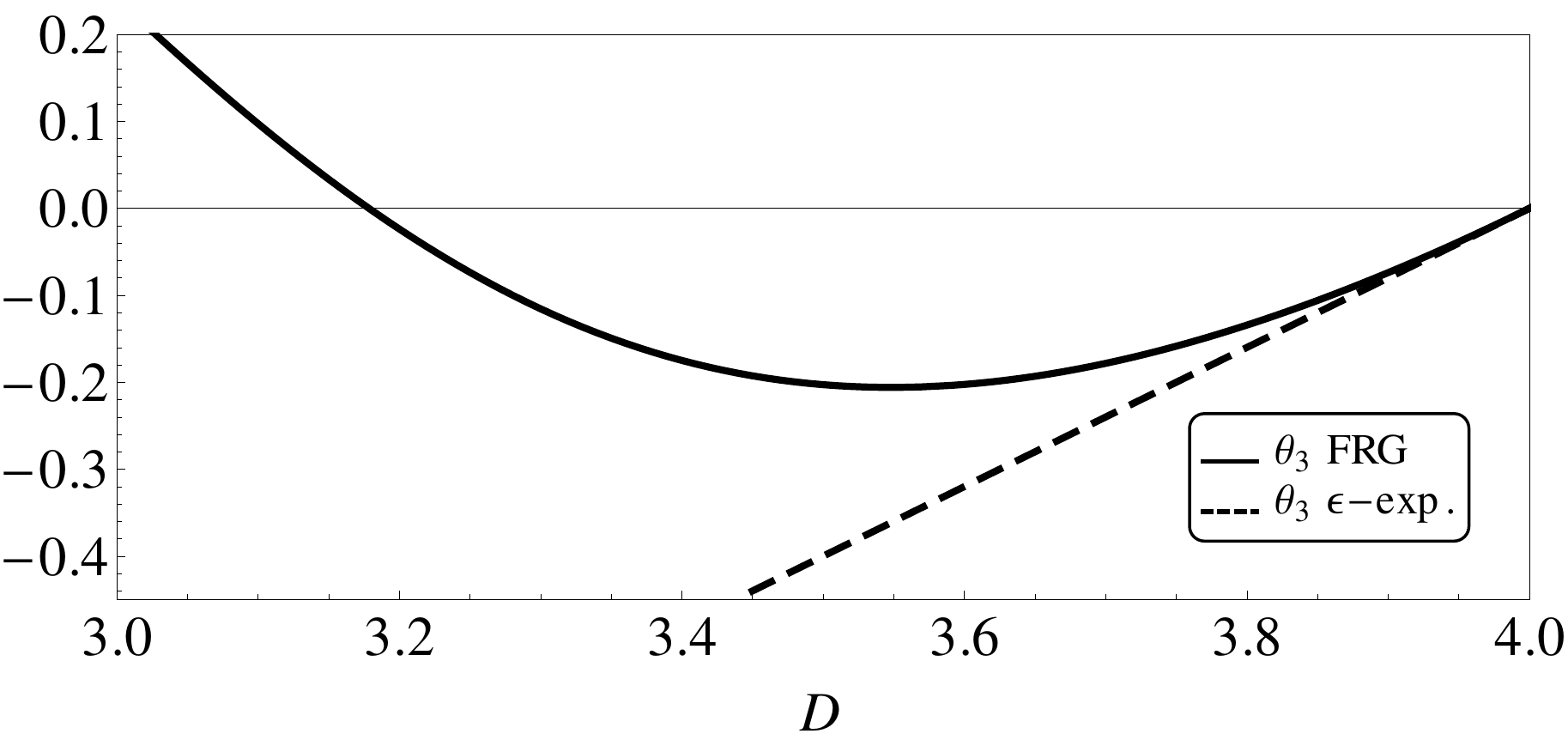}
 \caption{Three largest critical exponents $\theta_1$, $\theta_2$ (top), and $\theta_3$ (bottom) at the cH+I fixed point from FRG and $\epsilon$ expansion~\cite{classen2015} as function of the space-time dimensions, for $N_f=2$. The cH+I fixed point is stable close to four dimensions. In the FRG approach, however, it bends to positive values below $D=3.17$ and renders the cH+I fixed point unstable in $D=3$. }
\label{fig:dim}
\end{figure}

\begin{figure}[t!]
\centering
 \includegraphics[width=.9\columnwidth]{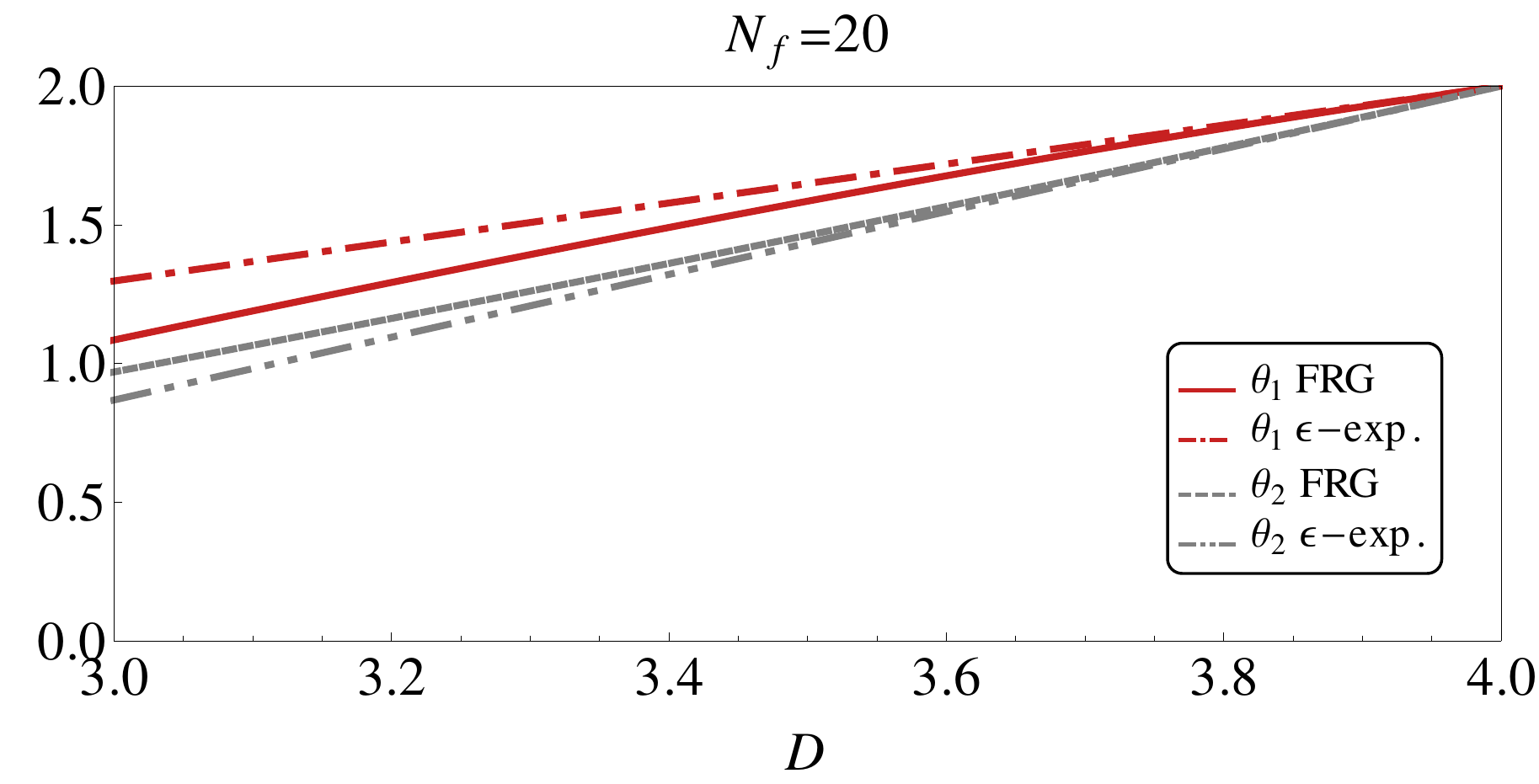}\\
 \includegraphics[width=.9\columnwidth]{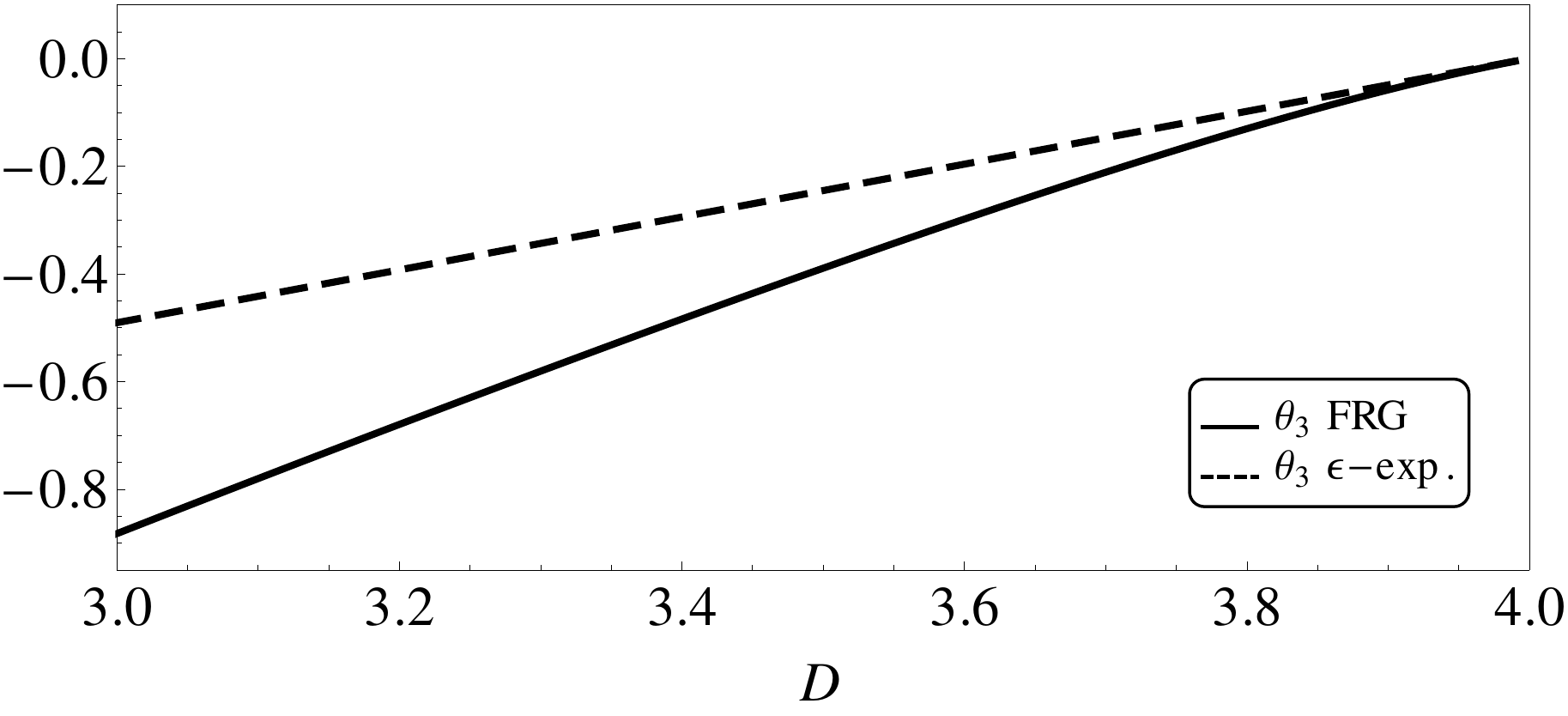}
 \caption{Three largest critical exponents $\theta_1$, $\theta_2$ (top), and $\theta_3$ (bottom) at large-$N_f$ fixed point from FRG and $\epsilon$ expansion~\cite{classen2015} as function of the space-time dimensions, for $N_f=20$. Here, the large-$N_f$ fixed point is stable in both approaches for all $3\leq D <4$.}
\label{fig:dim2}
\end{figure}

Using the FRG equations, we can directly evaluate the fixed points in arbitrary space-time dimensions $2<D<4$. 
This allows us to systematically compare to the fixed-point solutions of the $\epsilon$ expansion and track deviations when approaching $D=3$, i.e., for large values of $\epsilon$.
As explained in Sec.~\ref{subsec:FPs}, the study of the quantum multicritical point in first-order $\epsilon$ expansion has revealed two fixed points, which became stable at different ranges of the fermion flavor number $N_f$. 
For the graphene case, $N_f=2$, the $\epsilon$ expansion renders the cH+I fixed point stable, whereas a novel interacting fixed point that couples both chiral sectors of the theory became stable at large $N_f$. We will refer to this fixed point as ``large-$N_f$ fixed point'' in the following.
We can identify both fixed points within the FRG approach close to four space-time dimensions and then investigate their stability, as determined by the third-largest critical exponent $\theta_3$, as function of dimension $D$.
The evolution of all three largest critical exponents upon varying the dimension is depicted in Fig.~\ref{fig:dim} for the cH+I fixed point and in Fig.~\ref{fig:dim2} for the large-$N_f$ fixed point.

For the first two exponents the leading-order $\epsilon$-expansion results agree fairly well with the nonperturbative values of the FRG.
On the other hand, the decisive third exponent shows large deviations in three space-time dimensions. 
Regarding the cH+I FP, this effect can be traced back to the propagators in the loop contributions, which to first order in $\epsilon$, are accounted for only in the flow equations of the masses. 
However, for this fixed point $\theta_3$ is predominantly determined by $\beta_{g_\chi^2}$.
Dimensional analysis shows that $g_\chi^2$ scales like $(4-D)$, which is corrected by the loop contributions and the anomalous dimensions. 
These are much smaller in the FRG compared to the first order in $\epsilon$ due to the threshold effects of the propagators. 
Thus, while to first order in $\epsilon$ the loop contributions reduce $\theta_3$ below zero, the reduction is not as large in the nonperturbative setting. 
In contrast, main contributions to $\theta_1$ and $\theta_2$ come from the mass flow equations so that the $\epsilon$ expansion captures their behavior already quite well to first order. 
For larger values of $N_f$, loop contributions become less important and the critical exponents are mainly fixed by the canonical scaling. 
Here, the first order $\epsilon$ expansion underestimates the anomalous dimensions, so that the exponents tend faster to the final values of $\pm1$ within the FRG approach. 
The effect is larger for $\theta_3$ because the anomalous dimension enters it twice, both directly as well as through the derivative with respect to $g_\chi^2$.
Such significant quantitative improvement of the FRG approach as compared to the first-order $\epsilon$ expansion is well-known from the corresponding multicritical bosonic systems with $\mathrm O(N_1)\oplus \mathrm O(N_2)$ symmetry~\cite{eichhorn2013}, for which the true regions of stability of the different fixed points are by now well established~\cite{herbutbook}.

Eventually, the difference in $\theta_3$ leads to an important change of the stability analysis in three space-time dimensions for $N_f=2$ because the cH+I fixed point looses its stability at about $D = 3.17$. 
At this point it collides with a new fixed point that couples both sectors. 
However, the new fixed point is unphysical for $D < 3.17$ as it has a negative square of the Yukawa coupling $g_\chi^2<0$. 
Therefore, in $D=3$ no stable and physically admissible fixed point is found for the graphene case, i.e., all allowed fixed points have more than two relevant directions for $N_f=2$. 
%


\subsection{Dependence on $N_f$ in $D=3$}

In this section, we study the fixed-point structure as function of the fermion flavor number $N_f$ in three space-time dimensions. 
We find several regimes exhibiting the qualitative behavior known from first-order $\epsilon$ expansion: 
(1) For small $N_f$ the cH+I fixed point is stable, followed by a regime of (2) intermediate $N_f$ where no stable and physically admissible fixed point exists. 
(3) For large $N_f$, a novel FP with a coupling between the different sectors is stable.

\begin{figure}[t!]
\centering
 \includegraphics[width=0.9\columnwidth]{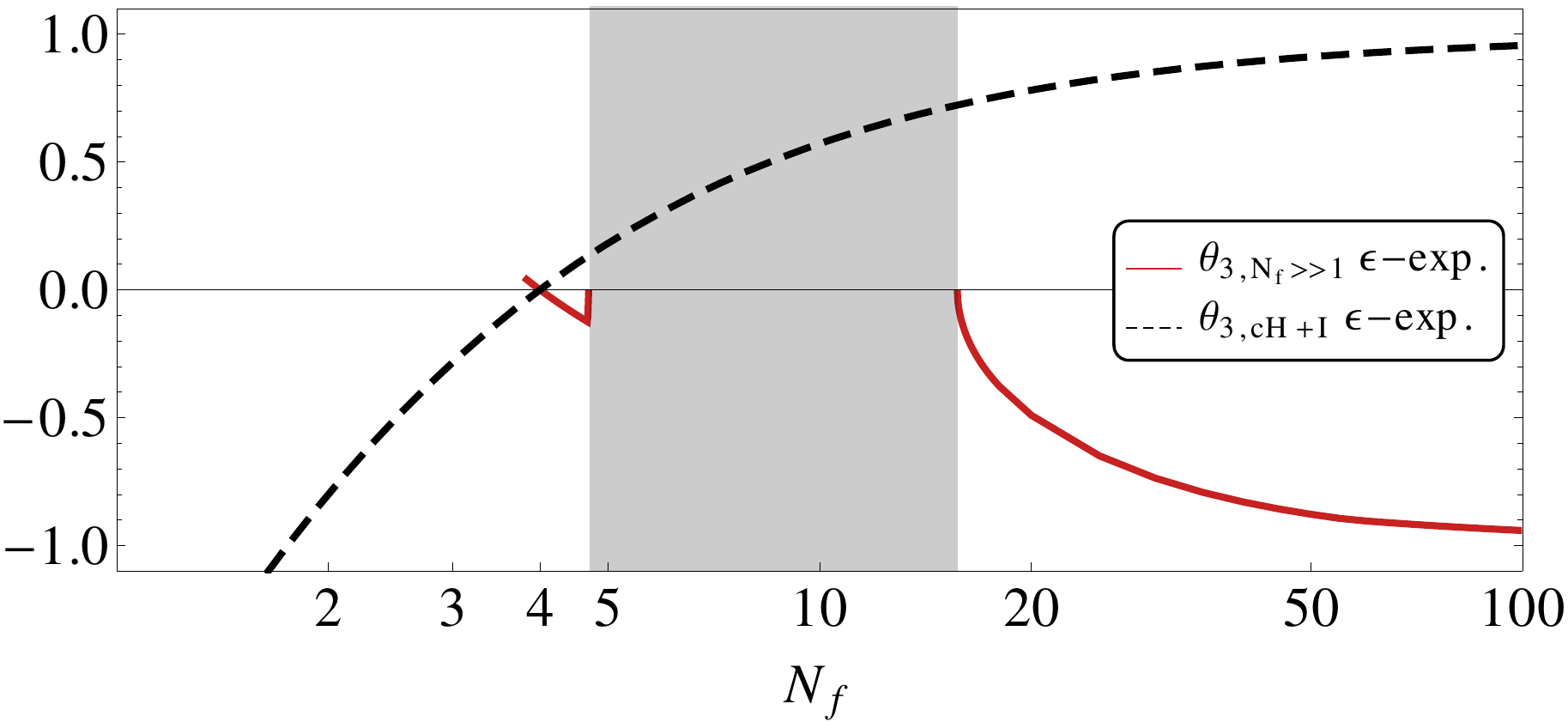}\\
 \includegraphics[width=0.9\columnwidth]{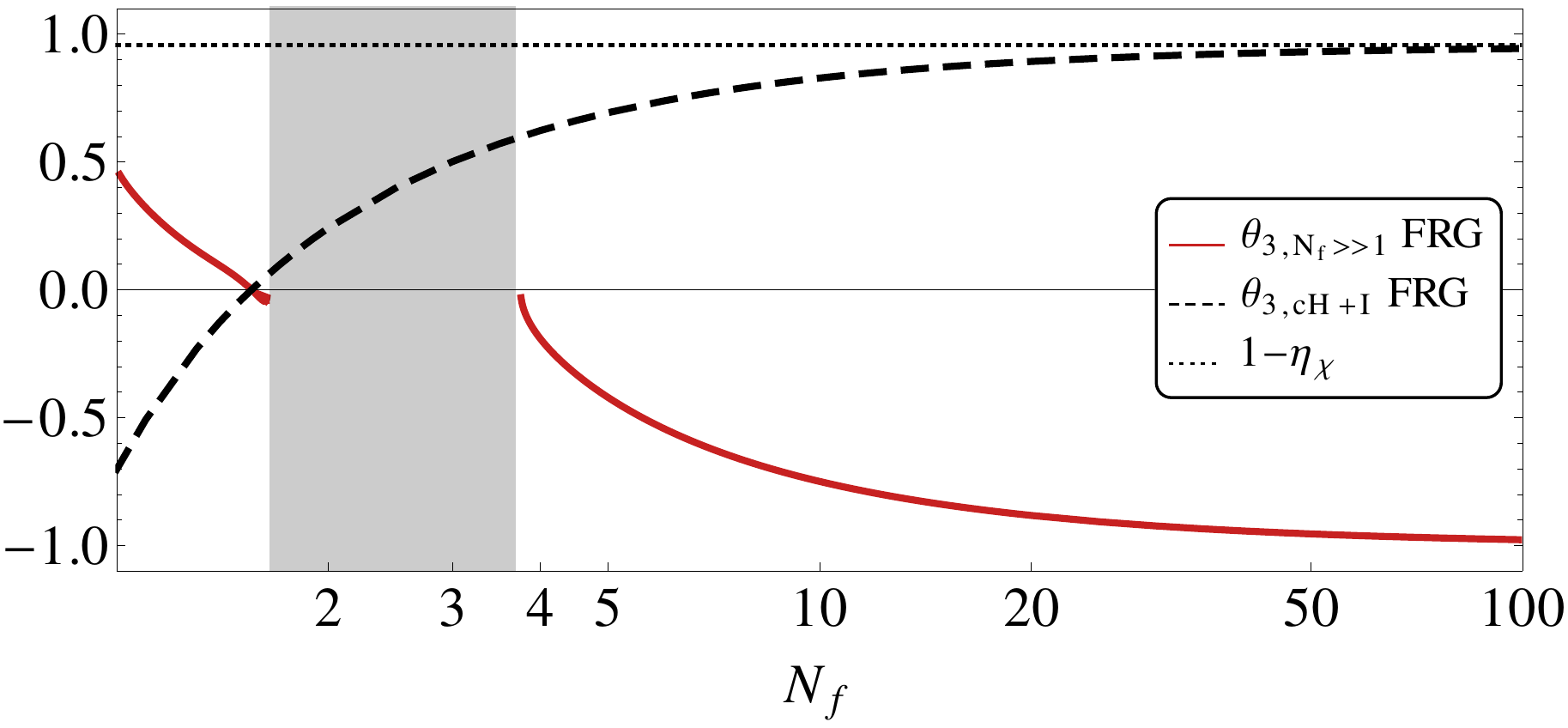}
 \caption{Third largest critical exponent $\theta_3$ as function of the fermion flavor number $N_f$ of cH+I fixed point (dashed/black) and large-$N_f$ fixed point (solid/red), from $\epsilon$ expansion~\cite{classen2015} (top) in comparison to FRG (bottom). We also show $1-\eta_\chi$ for the cH+I fixed point in the FRG (bottom), note the offset from 1 for large $N_f$.}
\label{fig:theta3Nf}
\end{figure}

On the other hand, the values of $N_f$ marking the borders change considerably when going from the first order $\epsilon$ expansion to the FRG approach.
We again investigate the sign of $\theta_3$ to analyze the stability and determine the critical flavor numbers for the different regimes. 
The result is depicted in Fig.~\ref{fig:theta3Nf}, where for comparison we also show the $\epsilon$-expansion results from Ref.~\cite{classen2015}. 
One can explicitly see that the qualitative behavior is the same within both approaches, but the different regimes are shifted and shrunk. 
Within the FRG, the first regime at small flavor numbers $N_f<1.6$ is characterized by the cH+I fixed point. 
At $N_f = 1.6$ it collides with another fixed point and looses stability. 
As pointed out in the previous section, the other fixed point is unphysical due to a negative $g_\chi^2$ for $N_f > 1.6$.  
Hence, for $N_f \in (1.6,3.6)$ no stable and physically admissible fixed point is found.
Finally, for $N_f>3.6$ the large-$N_f$ fixed point known from the $\epsilon$ expansion becomes stable. 
%
%

For numerical comparison, we give the three largest critical exponents and the anomalous dimensions for two examples of $N_f$, for which a stable fixed point is found, in Table~\ref{tab:stable}.
%
%
\begin{table}[t!]
\caption{\label{tab:stable} LPA8' results of the stable fixed point for $N_f=1$ and $N_f=20$ in $D=3$.}
\renewcommand{\arraystretch}{1.4}
\renewcommand{\tabcolsep}{4pt}
\begin{tabular}{c  c  c c c  c c c c}
\hline\hline
$N_f$ & stable FP & $\theta_1$ & $\theta_2$&$\theta_3$ & $\eta_\phi$ & $\eta_\chi$ & $\eta_\psi$\\
\hline
1 & cH+I & 1.564 &  0.558 & -0.703 & 1.003 &  0.044 &  0.207 \\
20 & large-$N_f$ & 1.085 &  0.969 & -0.883 & 0.980 & 0.913 &  0.010 \\
\hline\hline
\end{tabular}
\end{table}


\subsubsection{Large $N_f$ behavior}

To complete the stability analysis for the cH+I and the large-$N_f$ fixed points, we finally study the limit $N_f \to \infty$, for which we can calculate the exponents analytically.
To this end, we rescale the potential and the bosonic wave function renormalization by suitable factors of $N_f$, $U\rightarrow U/N_f, \ Z_{\chi/\phi}\rightarrow Z_{\chi/\phi}/N_f$, ensuring that $g_{\chi}$ and $g_\phi$ remain positive.
After this rescaling, the boson loops become of order $\mathcal{O}(1/N_f)$ and the flow equations to leading order read
\begin{align}\label{eqn:potential_largeNf}
\partial_t u&=-D u + (D-2 +\eta_\chi)\rho_\chi u^{(1,0)}\nonumber\\  
&+ (D-2 +\eta_\phi)\rho_\phi u^{(0,1)}\nonumber\\ 
&- 2\eck{I_\psi(\omega_\psi^+)+I_\psi(\omega_\psi^-)} + \mathcal{O}(1/N_f)\,,
\end{align}
and
\begin{align}
\partial_t g_{\chi/\phi}^2&=(D-4+\eta_{\chi/\phi})g_{\chi/\phi}^2+ \mathcal{O}(1/N_f)\label{eqn:YukawalargeNf}\,,\\
\eta_{\chi/\phi}&=\frac{32v_D}{D} m_4^{(F)}(0)g_{\chi/\phi}^2+\mathcal{O}(1/N_f)\,, \label{eq:anomalous-dimension-large-Nf}\\
\eta_\psi&=\mathcal{O}(1/N_f)\,,
\end{align}
with the threshold functions as given in Appendix~\ref{app:threshold} and $v_D=(\Gamma(D/2)2^{D+1}\pi^{d/2})^{-1}$. 

The problem becomes symmetric with respect to $\chi$ and ${\boldsymbol\phi}$ and can be solved exactly. In the sector of the Yukawa couplings, the fixed-point solutions are
\begin{align}
	 g_{\chi/\phi}^{*^2}&=0 & \text{or} && 
	 g_{\chi/\phi}^{*^2}&=\frac{(4-D)D}{32v_Dm_4^{(F)}(0)}\,.
\end{align}
Further, the partial differential equation for the effective potential is solved by
\begin{align}
u(\rho_\chi,\rho_\phi)&=\frac{8v_D}{D^2}\left[\frac{_2F_1\left(\frac{D}{2},\frac{D}{2};\frac{D+2}{2};\frac{\omega_-}{1+\omega_-}\right)}{\left(1+\omega_-\right)^{D/2}}-2\right.\nonumber\\
&\left. +\frac{_2F_1\left(\frac{D}{2},\frac{D}{2};\frac{D+2}{2};\frac{\omega_+}{1+\omega_+}\right)}{\left(1+\omega_+\right)^{D/2}}\right]+\rho _{\chi }^{D/2} c\Big(\frac{\rho _{\phi }}{\rho _{\chi }}\Big)\,,\nonumber\\
\text{with}\quad \omega_\pm&=\frac{4v_D}{D}\frac{(4-3 D)}{\left(D^2-6 d+8\right)}\left(\sqrt{\rho _{\phi }}\pm\sqrt{\rho _{\chi }}\right)^{-2}\,,
\end{align}
and $_2F_1(a,b;c;z)$ denotes the Gaussian hypergeometric function (see, e.g., \cite{gradshteyn}). For any smooth function $c(\rho_\phi/\rho_\chi)$ that depends only on the ratio of the invariants $\rho_\phi/\rho_\chi$ the corresponding $u(\rho_\chi,\rho_\phi)$ solves the fixed-point equation~(\ref{eqn:potential_largeNf}). We can restrict $c$ by the physical requirement that the effective potential should be bounded from below and finite for $\rho_\chi\rightarrow0$ and $\rho_\phi\rightarrow0$. Alternatively, it can be determined so that $u$ equals the large-$N_f$ limit of the Taylor-expanded effective potential.

Regarding the stability analysis, we find that in this limit the entries $\partial \beta_{g_\chi^2}/\partial g_\chi^2$ and $\partial \beta_{g_\phi^2}/\partial g_\phi^2$ in the stability matrix fix $\theta_3=\theta_4$.
As can be seen in Eq.~(\ref{eqn:YukawalargeNf}), $\theta_3$ is then determined by the canonical scaling only
\begin{align}\label{eqn:theta3_largeNf}
\theta_3 = -\frac{\partial \beta_{g_\chi^2}}{\partial g_\chi^2}=-\Big(D-4+\eta_\chi +g_\chi^2 \frac{\partial \eta_\chi}{\partial g_\chi^2} + \mathcal{O}\big(\frac{1}{N_f}\big)\Big).\nonumber
\end{align}
For the stable large-$N_f$ fixed point both Yukawa couplings are nonzero and uniquely determined by the requirements $\eta_\chi=\eta_\phi=4-D$. We furthermore have $g_\chi^2 \partial \eta_\chi/\partial g_\chi^2=\eta_\chi$ and $g_\phi^2 \partial \eta_\phi/\partial g_\phi^2=\eta_\phi$ [Eq.~\eqref{eq:anomalous-dimension-large-Nf}].
This requires that the third (and fourth) largest critical exponent $\theta_3$ must tend to minus one in $D=3$,
\begin{align}
\text{large-$N_f$ fixed point:}\qquad \lim_{N_f\to \infty}\theta_3 = -1\,.
\end{align}
To investigate the cH+I fixed point in the limit of $N_f \to \infty$, we scale only the Heisenberg sector with the factor $1/N_f$, since the Ising sector completely decouples and becomes purely bosonic. Again the third critical exponent is determined by $\beta_{g_{\chi}^2}$. But now only loops including $g_\phi^2$ are suppressed by $1/N_f$, such that $\theta_3$ is given by
\begin{align}
\theta_3=-\frac{\partial \beta_{g_\chi^2}}{\partial g_\chi^2}=&-\Bigg(D-4+\eta_\chi +g_\chi^2 \frac{\partial \eta_\chi}{\partial g_\chi^2}\nonumber \\
 &\quad +\frac{\partial}{\partial g_\chi^2}\mathcal{L}(g_\chi^4) + \mathcal{O}\left(\frac{1}{N_f}\right)\Bigg),
\end{align}
where $\mathcal{L}(g_\chi^4)$ denotes loops that are at least proportional to $g_\chi^4$. With $g_\chi^2=0$ this reduces to
\begin{align}
\text{cH+I:} \qquad \lim_{N_f \to \infty}\theta_3 = 1-\eta_\chi
\end{align}
in $D=3$ space-time dimensions. 
We therefore see that the third critical exponent for the cH+I fixed point computed within the FRG does {\it not} coincide with the one from $\epsilon$ expansion.
This is due to the contribution from the anomalous dimension in the Ising sector, which becomes nonvanishing only to second order in $\epsilon$. 
In our approximation we have $\eta_\chi=0.044$, such that $\eta_\chi \nrightarrow 1$ for large $N_f$.
The large-$N_f$ behavior of both fixed points is exhibited in Fig.~\ref{fig:theta3Nf}.


\subsection{Phase diagram}\label{sec:phasediag}

\begin{figure}[t!]
\centering
 \includegraphics[width=0.9\columnwidth]{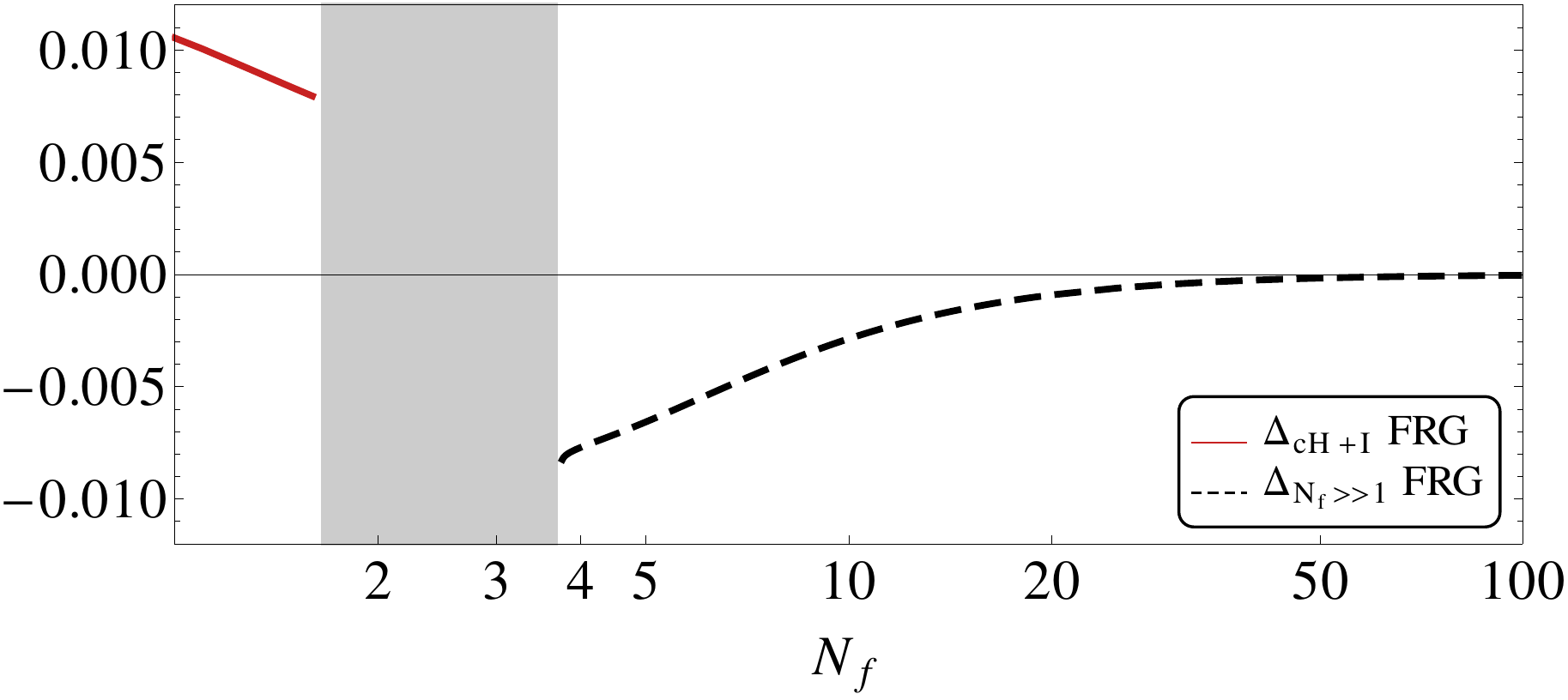}
 \caption{$\Delta=\lambda_{2,0}\lambda_{0,2}-\lambda_{1,1}^2$ at the stable fixed point for different fermion flavor numbers. Each $\lambda$ is scaled with $v_{D=3}$. For small $N_f<1.6$, $\Delta$ is positive at the stable (cH+I) fixed point, whereas for large $N_f>3.6$, $\Delta$ is negative at the stable (large-$N_f$) fixed point. For intermediate $N_f$ (gray shaded region), there is no stable and physically admissible fixed point.}
\label{fig:Delta}
\end{figure}

In the case that a stable, physical fixed point is found, the structure of the phase diagram near the multicritical point can be extracted from a simple criterion~\cite{liu1973,nelson1974,kivelson2001}.
We define
\begin{align}
\Delta=\lambda_{2,0}\lambda_{0,2}-\lambda_{1,1}^{2},
\end{align}
with $\lambda_{2,0}, \lambda_{0,2}$ and $\lambda_{1,1}$ being the corresponding expansion coefficients of the infrared effective potential.%
\footnote{Note that we use other conventions than in Ref.~\cite{classen2015} regarding the factors in the action Eq.~(\ref{eqn:actionB}), leading to a different factor in the definition of $\Delta$.}
For $\Delta>0$ the effective potential in the ordered phase is minimized when {\it both} sectors simultaneously develop a nonzero vacuum expectation value, corresponding to a phase of coexistence between the separate phases. 
The mixed phase cannot exist if $\Delta<0$, and the phase diagram exhibits bicritical behavior in this case. 
The criterion has been used in a variety of studies that investigate the competition between different order parameters~\cite{calabrese2003,eichhorn2013,eichhorn2014,classen2015}.
If we start the RG flow near the stable fixed point, it remains in its vicinity for a long RG ``time''.
In this way, the direct neighborhood of the multicritical point in the phase diagram should depend only on the properties of the effective potential {\it at the fixed point}.
To determine the behavior near the multicritical point, it then suffices to compute the value for $\Delta$ using the fixed-point values for the quartic couplings.
Eventually, of course, the system will flow away from the critical surface and the argument breaks down far from the multicritical point.
Fig.~\ref{fig:Delta} shows the value for $\Delta$ at the cH+I fixed point for $N_f<1.6$ and the large-$N_f$ fixed point for $N_f>3.6$.
$\Delta$ is positive at the cH+I, so that the phase diagram exhibits tetracritical behavior and a coexistence phase [situation I in Fig.~\ref{fig:phasediag}(a)] for small $N_f$.
On the other hand, $\Delta$ is negative at the large-$N_f$ fixed point, leading to  bicritical behavior with a first-order transition between the ordered states for large $N_f$. The transition across the multicritical point, however, remains continuous [situation III in Fig.~\ref{fig:phasediag}(a)].
This is in qualitative agreement with the previous result~\cite{classen2015}.

When a stable and physically admissible fixed point does {\it not} exist, the phase diagram close to the intersection of SM, SDW, and CDW is governed by a triple point, with {\it all} transitions in its vicinity appearing first order [situation II in Fig.~\ref{fig:phasediag}(a)].
As an important observation within the FRG approach, we in fact find this situation to be realized for the physical case of graphene, $N_f=2$ (gray shaded region in Fig.~\ref{fig:Delta}).
%


\section{Conclusions}\label{sec:conclusion}

We have studied a multicritical point in the phase diagram of electrons on the honeycomb lattice using an effective field theory as low-energy theory of an extended Hubbard model with onsite and nearest-neighbor interaction. 
Our theory accounts for the universal behavior in the regime where the semimetallic phase, the charge density wave phase, and the spin density wave phase meet.
Within a nonperturbative FRG approach we were able to investigate the dependence on space-time dimension $2<D<4$ and flavor number $N_f$, thereby extending a previous study close to four space-time dimensions~\cite{classen2015}. 

We have calculated the fixed-point structure and its stability ranges to describe the competition of the different phases. 
This enables us to determine the nature of the transition lines and the possibility of a CDW-SDW coexistence phase, as function of the number of fermion flavors.
Besides, we provide a quantitative description of the critical behavior in the cases when the transitions are continuous.
We have followed the two fixed points that are stable (at different ranges of $N_f$) from the upper critical space-time dimension $D=4$ down to $D=3$. While our results agree near $D=4$ both qualitatively and quantitatively with the $\epsilon$-expansion results~\cite{classen2015}, we have found significant quantitative changes in the decisive third critical exponent and anomalous dimensions in $D=3$.
This leads to modified stability ranges, although the qualitative picture remains the same as in the $\epsilon$ expansion. 
The borders of the different stability regimes are determined by the collision of fixed points. 
They move in theory space as function of dimension and fermion flavor number and exchange stability when they meet.

Explicitly, we have found three different regimes for varying fermion flavor number.
For small number of flavors the cH+I fixed point determines the physical properties at the multicritical point. 
In other words, the semimetal-to-antiferromagnet transition determines also the universal behavior at the multicritical point. Beyond this point, our results suggests a mixed phase in which both SDW and CDW orders coexist (tetracritical point).
For large $N_f$ a new fixed point of the coupled system emerges and becomes stable. 
The transition between both ordered states now is first order, whereas directly at the multicritical point the transition is continuous and defines a novel universality class (bicritical point), in agreement with the large-$N_f$ calculations within the fermionic description~\cite{herbut2006}.
The graphene case is placed in a third regime that occurs for intermediate $N_f$. 
Here we do not find any stable fixed point, leading us to the prediction that a triple point appears with first-order transitions only.
While in our description the number of fermion flavors has been introduced merely as a theoretical control parameter, a similar deformation of the honeycomb lattice system should be relevant for the novel systems with a large number of Dirac cones \cite{miert2015}.

We have employed the FRG in terms of a local potential approximation within the derivative expansion including anomalous dimensions. 
As a crosscheck, we have compared our results with known limits from, on the one hand, the separate universality classes~\cite{janssen2014}, and, on the other hand, the $\epsilon$-expansion results near the upper critical dimension~\cite{classen2015}. 
In both limits we find perfect agreement, as it should be.
We have also compared our predictions for $D=3$ with various literature results.
As shown in \cite{janssen2014}, the FRG critical exponents in the Gross-Neveu-Yukawa model of the separate transitions also become exact close to the lower critical dimension of the corresponding, purely fermionic Gross-Neveu model, and  the FRG interpolates continuously in between both exact limits.
We expect this also to hold for the additional critical exponents arising from the coupling of both the chiral Ising and the chiral Heisenberg sectors.
Concerning the convergence of our polynomial truncation we have verfied the convergence of the critical exponents upon inclusion of higher polynomial orders within our potential expansion, see Appendix~\ref{app:conv}. 
However, in order to resolve the discrepancy between FRG and Monte Carlo results~\cite{sorella2012, assaad2013, toldin2015} for the critical exponents of the chiral Heisenberg universality class it might be needed to go beyond our approximation.
This could be done, for instance, by accounting for field-dependent Yukawa couplings \cite{vacca2015} and/or field-dependent wave function renormalizations.
In addition it would be interesting to compute the global behavior of the fixed-point potential at finite $N_f$, in particular in light of the first-order phase transitions we predict.
This could be done, for instance, along the lines suggested in Ref.~\cite{borchardt2015}.
%


\acknowledgments

We acknowledge discussions with I. Boettcher, J. Borchardt, A. Eichhorn, H. Gies, B. Knorr, F. Rennecke, L. von Smekal, and S. Wessel. L.C.\ acknowledges support by the Studienstiftung des deutschen Volkes. M.M.S.\ is supported by Grant No.\ ERC-AdG-290623. I.F.H.\ and L.J.\ were supported by the NSERC of Canada. L.J.\ was also supported by the DFG under Grant Nos.  JA\,2306/1-1, JA\,2306/3-1, and SFB1143.


\appendix

\begin{widetext}


\section{Flow equations for the Yukawa couplings}\label{app:betafcts}

Here, we explicitly display the full analytical expressions for the flow of the two Yukawa couplings in terms of threshold functions, which contain the loop integration and regulator dependence. 
The definitions for the threshold functions and explicit expressions for the case of particular regulator choices can be found in Appendix~\ref{app:threshold}. 
Note that the threshold functions are defined such that they also depend on the Yukawa couplings. Then they reduce to familiar expressions in the uncoupled limit ($\omega_{\phi\chi}=0$).
First, we display the flow equation of the Yukawa coupling to the Ising field $\chi$:
\begin{align}
\partial_t g_{\chi}^2&=(D-4 +\eta_{\chi}+2\eta_\psi)g_\chi^2 \nonumber\\ 
&+4v_D\sum_{\sigma=\pm 1}\left\{ g_\chi^3 l_{(11)R_\chi}^{(FB),\sigma}(g_\chi,g_\phi;\omega_\chi,\omega_\phi,\omega_{\phi\chi},\omega_\psi^\sigma)+g_\chi^2 g_\phi\left[ l_{(11)R_\phi}^{(FB),\sigma}(g_\phi,g_\chi;\omega_\phi,\omega_\chi,\omega_{\phi\chi},\omega_\psi^\sigma)  + 2 g_\phi l_{(11)G_\phi}^{(FB)}(u^{(0,1)},\omega_\psi^\sigma)\right] \right. \nonumber\\
&-2(\sqrt{2\kappa_\chi}g_\chi+\sigma\sqrt{2\kappa_\phi}g_\phi)^2\left[  g_\chi^3  l_{(21)R_\chi}^{(FB),\sigma}(g_\chi,g_\phi;\omega_\chi,\omega_\phi,\omega_{\phi\chi},\omega_\psi^\sigma)\right. \nonumber\\
&\left.+ g_\chi^2g_\phi \left(  l_{(21)R_\phi}^{(FB),\sigma}(g_\phi,g_\chi;\omega_\phi,\omega_\chi,\omega_{\phi\chi},\omega_\psi^\sigma)+ 2 g_\phi l_{(21)G_\phi}^{(FB)}(u^{(0,1)},\omega_\psi^\sigma) \right)\right] \nonumber\\
&-g_\chi(\sqrt{2\kappa_\chi}g_\chi+\sigma\sqrt{2\kappa_\phi}g_\phi) \left[  \omega_{\chi\chi\chi}  l_{(12)R_\chi}^{(FB),\sigma}(g_\chi,g_\phi;\omega_\chi,\omega_\phi,\omega_{\phi\chi},\omega_\psi^\sigma) + \omega_{\chi\phi\phi} l_{(12)R_\phi}^{(FB),\sigma}(g_\phi,g_\chi;\omega_\phi,\omega_\chi,\omega_{\phi\chi},\omega_\psi^\sigma)\right.  \nonumber\\
&\left.\left. + 2g_\phi^2\sqrt{\kappa_\chi}u^{(1,1)} l_{(12)G_\phi}^{(FB)}(u^{(0,1)},\omega_\psi^\sigma) -2\omega_{\chi\chi\phi}l_{(111)R_\chi R_\phi}^{(FBB),\sigma}(g_\chi,g_\phi;\omega_\chi,\omega_\phi,\omega_{\phi\chi},\omega_\psi^\sigma)\right] \right\}\,.
\end{align}
Analogously, the Yukawa coupling to the Heisenberg field ${\boldsymbol\phi}$ is given by
\begin{align}
\partial_t g_\phi^2&= (D-4 +\eta_{\phi}+2\eta_\psi)g_\phi^2 \nonumber\\ 
&+4v_D\sum_{\sigma=\pm 1}\left\{ -g_\phi^3 l_{(11)R_\phi}^{(FB),\sigma}(g_\phi,g_\chi;\omega_\phi,\omega_\chi,\omega_{\phi\chi},\omega_\psi^\sigma) +g_\phi^2 g_\chi l_{(11)R_\chi}^{(FB),\sigma}(g_\chi,g_\phi;\omega_\chi,\omega_\phi,\omega_{\phi\chi},\omega_\psi^\sigma) \right.\nonumber\\
&-4g_\phi^2 g_\chi^2 \kappa_\chi\left[g_\chi l_{(111)R_\chi}^{(FFB),\sigma}(g_\chi,g_\phi;\omega_\chi,\omega_\phi,\omega_{\phi\chi},\omega_\psi^+,\omega_\psi^-) -g_\phi l_{(111)R_\phi}^{(FFB),\sigma}(g_\phi,g_\chi;\omega_\phi,\omega_\chi,\omega_{\phi\chi},\omega_\psi^+,\omega_\psi^-) \right] \nonumber\\
& -2g_\phi^2u^{(1,1)}\sqrt{2\kappa_\chi}(\sqrt{2\kappa_\chi}g_\chi+\sigma\sqrt{2\kappa_\phi}g_\phi)
l_{(111)G_\phi R_\chi}^{(FBB),\sigma}(g_\chi,g_\phi;u^{(0,1)},\omega_\chi,\omega_\phi,\omega_{\phi\chi},\omega_\psi^\sigma) \nonumber\\
&\left. -2g_\phi^2u^{(0,2)}\sqrt{2\kappa_\phi}(\sqrt{2\kappa_\phi}g_\phi+\sigma\sqrt{2\kappa_\chi}g_\chi) l_{(111)G_\phi R_\phi}^{(FBB),\sigma}(g_\phi,g_\chi;u^{(0,1)},\omega_\phi,\omega_\chi,\omega_{\phi\chi},\omega_\psi^\sigma) \right\}\,.
\end{align}
We have abbreviated the volume factor by $v_D=\frac{1}{4}\text{vol}(S^{D-1})/(2\pi)^D=(\Gamma(D/2)2^{D+1}\pi^{d/2})^{-1}$. The threshold functions are listed in Appendix~\ref{app:threshold} and we have defined
\begin{align}
\omega_{\chi\chi\chi}=\sqrt{2\kappa_\chi}(3u^{(2,0)}+2\kappa_\chi u^{(3,0)})\,,\quad
\omega_{\chi\chi\phi}=\sqrt{2\kappa_\phi}(u^{(1,1)}+2\kappa_\chi u^{(2,1)})\,,\quad
\omega_{\chi\phi\phi}=\sqrt{2\kappa_\chi}(u^{(1,1)}+2\kappa_\phi u^{(1,2)}).
\end{align}
$\omega_\phi$, $\omega_\chi$, $\omega_{\phi\chi}$, and $\omega_\psi^{\pm}$ are defined in the main text, see Sec.~\ref{sec:frgflow}.
%

\section{Anomalous dimensions}\label{app:anom}

The expressions for the anomalous dimensions can also be given as a closed algebraic system of equations.
Note that---although not exlicitly displayed---the threshold functions also depend on the anomalous dimensions. First, we list the expressions for the anomalous dimension of the Heisenberg field $\eta_\phi$ and the one for the Dirac fermions $\eta_\psi$,
\begin{align}
\eta_\phi&=\frac{4v_D}{D}\Big\{ 2\sqrt{2\kappa_\chi}u^{(1,1)}m_{(22)R_\chi G_\phi}^{(B)D,\sigma}(\sqrt{2\kappa_\chi}u^{(1,1)},\sqrt{2\kappa_\phi}u^{(0,2)};\omega_\chi,\omega_\phi,\omega_{\phi\chi},u^{(0,1)})\nonumber \\
&\quad+2 \sqrt{2\kappa_\phi}u^{(0,2)}m_{(22)R_\phi G_\phi}^{(B)D,\sigma}(\sqrt{2\kappa_\phi}u^{(0,2)},\sqrt{2\kappa_\chi}u^{(1,1)};\omega_\phi,\omega_\chi,\omega_{\phi\chi},u^{(0,1)})\nonumber\\
&\quad+2N_fd_\gamma g_\phi^2\left[  m_{(22)}^{(F)}(\omega_\psi^+,\omega_\psi^-) - (2\kappa_\chi g_\chi^2- 2\kappa_\phi g_\phi^2)m_{(11)}^{(F)}(\omega_\psi^+,\omega_\psi^-) \right] \Big\}\,,\\
\eta_\psi&=\frac{4v_D}{D}\sum_\sigma\Big\{ g_\chi m_{(12)R_\chi}^{(FB)D,\sigma}(g_\chi,g_\phi;\omega_\chi,\omega_\phi,\omega_{\phi\chi},\omega_\psi^\sigma) + 2g_\phi^2 m_{(12)G_\phi}^{(FB)}(u^{(0,1)},\omega_\psi^\sigma)  \nonumber\\
&\quad+ g_\phi m_{(12)R_\phi}^{(FB)D,\sigma}(g_\phi,g_\chi;\omega_\phi,\omega_\chi,\omega_{\phi\chi},\omega_\psi^\sigma) \Big\}\,.
\end{align}
The expression for the anomalous dimension of the Ising field $\eta_\chi$ depends on the projection prescription.
Here, we distinguish between the projections onto the radial mode or onto an auxiliary Goldstone mode (see Appendix~\ref{app:projection} for technical details).
Both ways are given in the following and are denoted as $\eta_{\chi,R}$ and $\eta_{\chi,G}$, respectively
\begin{align}
\eta_{\chi,R}&=\frac{4v_D}{D}\Big\{ m_{(40)R_\chi}^{(B)D,\sigma}(\omega_{\chi\chi\chi},\omega_{\chi\chi\phi};\omega_\chi,\omega_\phi,\omega_{\phi\chi}) + m_{(40)R_\phi}^{(B)D,\sigma}(\omega_{\chi\phi\phi},\omega_{\chi\chi\phi};\omega_\phi,\omega_\chi,\omega_{\phi\chi})\nonumber\\
&\quad+2m_{(22)R_\chi R_\phi}^{(B)D,\sigma}(\omega_{\chi\chi\phi},\omega_{\chi\phi\phi},\omega_{\chi\chi\chi};\omega_\chi,\omega_\phi,\omega_{\phi\chi})+4\kappa_\chi ( u^{(1,1)})^2 m_{(40)G_\phi}^{(B)D}(u^{(0,1)})  \nonumber\\
&\quad+ N_fd_\gamma g_\chi^2\sum_{\sigma}\left[ m_4^{(F)}(\omega_\psi^\sigma)-(\sqrt{2\kappa_\chi}g_\chi+\sigma\sqrt{2\kappa_\phi}g_\phi)^2m_2^{(F)}(\omega_\psi^\sigma)\right] \Big\}\,,\\
\eta_{\chi,G}&=\frac{4v_D}{D}\Big\{ 2\sqrt{2\kappa_\chi}u^{(2,0)}m_{(22)R_\chi G_\chi}^{(B)D,\sigma}(\sqrt{2\kappa_\chi}u^{(2,0)},\sqrt{\kappa_\phi}u^{(1,1)};\omega_\chi,\omega_\phi,\omega_{\phi\chi},u^{(1,0)})\nonumber\\ &\quad+ 2\sqrt{2\kappa_\phi}m_{(2,2)R_\phi G_\chi}^{(B)D,\sigma}(\sqrt{2\kappa_\phi}u^{(1,1)}\sqrt{2\kappa_\chi}u^{2,0};\omega_\phi,\omega_\chi,\omega_{\phi\chi},u^{(1,0)})\nonumber\\
&\quad+ N_fd_\gamma g_\chi^2\sum_{\sigma}\left[ m_4^{(F)}(\omega_\psi^\sigma)-(\sqrt{2\kappa_\chi}g_\chi+\sigma\sqrt{2\kappa_\phi}g_\phi)^2m_2^{(F)}(\omega_\psi^\sigma)\right] \Big\}\,.
\end{align}
For our results as displayed in the main text, we use the latter definition, which in the case of the purely bosonic systems is known to yield more accurate results, cf.~Appendix~\ref{app:projection}.


\section{Projection prescriptions for anomalous dimensions}\label{app:projection}

Let us consider an $N$-component bosonic field ${\boldsymbol\phi}$, which we divide into one radial and $N-1$ Goldstone modes ${\boldsymbol\phi}=(\phi_R,{\boldsymbol\phi}_G)$. 
In the following discussion, we will suppress the bosonic potential, since it plays no role for the argument.
Therefore, we use the truncation
\begin{align}
	\Gamma=\int_x \left[\frac{1}{2}Z(\partial_\mu{\boldsymbol\phi})^2 + \frac{1}{4}Y(\partial_\mu \rho)^2\right],
\end{align}
where we account for the first correction to the kinetic term in a derivative expansion and have defined $\rho=\frac{1}{2}{\boldsymbol\phi}^2$. 
The second functional derivative of $\Gamma$ gives
\begin{align}
\frac{\delta^2\Gamma}{\delta\phi_i\delta\phi_{j}}= 
- Z\delta_{ij}\partial_\mu^2- \frac{1}{2}Y\left(\delta_{ij} \partial_\mu^2\rho + \phi_j\partial_\mu^2\phi_i\right)\,,
\end{align}
where the derivative has to be understood as momentum squared in momentum representation. 
This shows that the projection $\left.\frac{\partial}{\partial p^2}\frac{\delta^2}{\delta \phi_R^2}\partial_t \Gamma\right|_0$ yields an expression proportional to $\partial_t Z +2\kappa\partial_t Y$ with $\kappa=\frac{1}{2}\phi_R^2$. 
On the other hand $\left.\frac{\partial}{\partial p^2}\frac{\delta^2}{\delta {\boldsymbol\phi}_G^2}\partial_t \Gamma\right|_0$ really projects onto $\partial_tZ$.

To circumvent the projection onto the radial mode in the case of the one-component Ising field, we choose a truncation with $N$ copies of it 
\begin{align}
	\Gamma_k&=\int d^Dx~\Big(
	Z_{\Psi,k}\bar\Psi(\mathds{1}_2\otimes \gamma_\mu)\partial_\mu\Psi-\frac{1}{2}Z_{\chi,k}\chi_a\partial_\mu^2\chi_a -\frac{1}{2}Z_{\phi,k}{\boldsymbol\phi}\partial_\mu^2{{\boldsymbol\phi}} \nonumber \\
& \quad+ \bar{g}_{\chi, k}\left(\sum_a\chi_a\right)\bar\Psi (\mathds{1}_2\otimes\mathds{1}_4)\Psi +\bar{g}_{\phi, k}{\boldsymbol\phi}\bar\Psi ({\boldsymbol{\sigma}}\otimes\mathds{1}_4)\Psi + U_k(\bar\rho_\chi,\bar\rho_\phi) \Big).
\end{align}
The field is divided into its vacuum expectation value and the fluctuations $\chi=(\chi_R+\Delta\chi_{R,1},\Delta\chi_{G,1},\ldots,\Delta\chi_{G,N-1})$ and we project onto one of the Goldstone modes according to Eq.~(\ref{eqn:etachi}). In the end of the procedure we again set $N=1$.


\section{Threshold functions}\label{app:threshold}
\label{app:thresh}

The threshold functions encode the loop integrations and the regulator dependences. Here, for convenience, we define the threshold functions such that a dependence on the Yukawa couplings is included as well. In the following, we will first list the expressions for general regulator functions.


\subsection{General expressions}

We write the regulators in the form $R^{(B)}_{i,k}(q)=Z_{i,k}q^2r_{i,k}(q^2)$ with $i\in\{\chi,{\boldsymbol\phi}\}$ and $R_k^{(F)}=iZ_{\psi,k} q_\mu\left(\mathbbm{1}_2\otimes\gamma_\mu\right)r_{\psi,k}(q)$. Further, we define $p_i(q)=q^2(1+r_{i,k}(q))$ and $p_F(q)=q^2(1+r_{\psi,k}(q))^2$. The derivative acting only on the regulator's $t$-dependence then reads
\begin{align}
\tilde\partial_t&=\sum_{\Phi\in\{\chi,\phi,\psi\}}\int dq~2q\frac{1}{Z_{\Phi,k}}\partial_t \eck{Z_{\Phi,k} r_{\Phi,k}(q)} \frac{\delta}{\delta r_{\Phi,k}(q)}\,.
\end{align}
For the threshold functions appearing in the flow equations for the Yukawa couplings, we then obtain
\begin{align*}
l_{(nm)R_\varphi}^{(FB),\sigma}(g_\varphi,g_\theta;\omega_\varphi,\omega_\theta,\omega_{\varphi\theta},\omega_\psi^\sigma)&=-\frac{1}{4v_D}k^{4-D}k^{2(n+m-2)}\tilde{\partial}_t\int_q\frac{(g_\varphi(p_\theta+k^2\omega_\theta)-\sigma g_\theta k^2\omega_{\varphi\theta})^m}{(p_F+k^2\omega_\psi^\sigma)^n((p_\varphi+k^2\omega_\varphi)(p_\theta+k^2\omega_\theta)-k^4\omega_{\varphi\theta}^2)^m}\,, \\
l_{(nm)G_\phi}^{(FB)}(\omega_\phi,\omega_\psi)&=-\frac{1}{4v_D}k^{4-D}k^{2(n+m-2)}\tilde{\partial}_t\int_q\frac{1}{(p_F+k^2\omega_\psi^\sigma)^n(p_\phi+k^2\omega_\phi)^m}\,,
\end{align*}
and
\begin{align*}
l_{(111)R_{\varphi}R_{\theta}}^{(FBB),\sigma}(g_{\varphi},g_{\theta};\omega_{\varphi},\omega_{\theta},\omega_{\varphi\theta},\omega_\psi^\sigma)&=-\frac{1}{4v_D}k^{6-D}\tilde{\partial}_t\int_q\frac{(g_{\varphi}(p_{\theta}+k^2\omega_{\theta})-\sigma g_{\theta}k^2\omega_{\varphi\theta})(g_{\theta}(p_{\varphi}+k^2\omega_{\varphi})-\sigma g_{\varphi}k^2\omega_{\varphi\theta})}{(p_F+k^2\omega_\psi^\sigma)((p_{\varphi}+k^2\omega_{\varphi})(p_{\theta}+k^2\omega_{\theta})-k^4\omega_{\varphi\theta}^2)^2}\,,\\
l_{(111)G_{\phi}R_{\varphi}}^{(FBB),\sigma}(g_{\varphi},g_{\theta};\omega_\phi,\omega_{\varphi},\omega_{\theta},\omega_{\varphi\theta},\omega_\psi^\sigma)&=-\frac{1}{4v_D}k^{6-D}\tilde{\partial}_t\int_q\frac{g_{\varphi}(p_{\theta}+k^2\omega_{\theta})-\sigma g_{\theta}k^2\omega_{\varphi\theta}}{(p_F+k^2\omega_\psi^\sigma)(p_\phi+k^2\omega_\phi)((p_{\varphi}+k^2\omega_{\varphi})(p_{\theta}+k^2\omega_{\theta})-k^4\omega_{\varphi\theta}^2)}\,,\\
l_{(111)R_{\varphi}}^{(FFB),\sigma}(g_{\varphi},g_{\theta};\omega_{\varphi},\omega_{\theta},\omega_{\varphi\theta},\omega_{\psi1},\omega_{\psi2})&=-\frac{1}{4v_D}k^{6-D}\tilde{\partial}_t\int_q\frac{g_{\varphi}(p_{\theta}+k^2\omega_{\theta})-\sigma g_{\theta}k^2\omega_{\varphi\theta}}{(p_F+k^2\omega_{\psi1})(p_F+k^2\omega_{\psi2})((p_{\varphi}+k^2\omega_{\varphi})(p_{\theta}+k^2\omega_{\theta})-k^4\omega_{\varphi\theta}^2)}\,.
\end{align*}
The threshold functions appearing in the anomalous dimensions can be distinguished by their internal lines. First, we have threshold functions corresponding to purely fermionic loops

\begin{align*}
m_4^{(F)D}(\omega)&=-\frac{1}{4v_D}k^{4-D}\tilde{\partial}_t\int_q q^4\left( \frac{\partial}{\partial q^2}\frac{1+r_\psi}{p_F + k^2 \omega} \right)^2\,, \\
m_{(22)}^{(F)D}(\omega_1,\omega_2)&=-\frac{1}{4v_D}k^{4-D}\tilde{\partial}_t\int_q q^4\left( \frac{\partial}{\partial q^2}\frac{1+r_\psi}{p_F + k^2 \omega_1} \right)\left( \frac{\partial}{\partial q^2}\frac{1+r_\psi}{p_F + k^2 \omega_2} \right)\,,\\
m_2^{(F)D}(\omega)&=-\frac{1}{4v_D}k^{6-D}\tilde{\partial}_t\int_q q^2\left( \frac{\partial}{\partial q^2}\frac{1}{p_F + k^2 \omega} \right)^2\,,\\
m_{(11)}^{(F)D}(\omega)&=-\frac{1}{4v_D}k^{6-D}\tilde{\partial}_t\int_q q^2\left( \frac{\partial}{\partial q^2}\frac{1}{p_F + k^2 \omega_1} \right)\left( \frac{\partial}{\partial q^2}\frac{1}{p_F + k^2 \omega_2} \right)\,.
\end{align*}
Then, we have purely bosonic contributions
\begin{align*}
m_{(22)R_\varphi R_\theta}^{(B)D}(v_1,v_2,v_3;\omega_\varphi,\omega_\theta,\omega_{\varphi\theta})&=-\frac{1}{4v_D}k^{6-D}\tilde{\partial}_t\int_q q^2\left(  \frac{\partial}{\partial q^2}\frac{(p_\varphi-k^2\omega_\varphi)v_1-k^2\omega_{\varphi\theta}v_2}{(p_\varphi+k^2\omega_\varphi)(p_\theta-k^2\omega_\theta)-k^4\omega_{\varphi\theta}^2} \right)\,,\\
&\hspace{3cm}\times\left(  \frac{\partial}{\partial q^2}\frac{(p_\theta-k^2\omega_\theta)v_1-k^2\omega_{\varphi\theta}v_3}{(p_\varphi+k^2\omega_\varphi)(p_\theta-k^2\omega_\theta)-k^4\omega_{\varphi\theta}^2} \right)\,,\\
m_{(22)R_\varphi G_\phi}^{(B)D}(v_1,v_2;\omega_\varphi,\omega_\theta,\omega_{\varphi\theta},\omega_\phi)&=-\frac{1}{4v_D}k^{6-D}\tilde{\partial}_t\int_q q^2\left(  \frac{\partial}{\partial q^2} \frac{(p_\theta+k^2\omega_\theta)v_1-k^2\omega_{\varphi\theta}v_2}{(p_\varphi+k^2\omega_\varphi)(p_\theta-k^2\omega_\theta)-k^4\omega_{\varphi\theta}^2}\right)\left(\frac{\partial}{\partial q^2}\frac{1}{p_\phi+k^2\omega_\phi}\right)\,,\\
m_{(40)R_\varphi}^{(B)D,\sigma}(v_1,v_2;\omega_\varphi,\omega_\theta,\omega_{\varphi\theta})&=-\frac{1}{4v_D}k^{6-D}\tilde{\partial}_t\int_q q^2\left(\frac{\partial}{\partial q^2}\frac{(p_\theta+k^2\omega_\theta)v_1-k^2\omega_{\varphi\theta}v_2}{(p_\varphi+k^2\omega_\varphi)(p_\theta+k^2\omega_\theta)-k^4\omega_{\varphi\theta}^2}\right)^2\,,\\
m_{(40)G_\phi}^{(B)D}(\omega_\phi)&=-\frac{1}{4v_D}k^{6-D}\tilde{\partial}_t\int_q q^2\left(  \frac{\partial}{\partial q^2}\frac{1}{p_\phi+k^2\omega_\phi} \right)^2\,.
\end{align*}
Finally, there are also threshold functions with mixed fermion-boson loops
\begin{align*}
m_{(12)G_\varphi}^{(FB)D}(\omega_{\varphi},\omega_\psi)&=-\frac{1}{4v_D}k^{4-D}\tilde{\partial}_t\int_q q^2\frac{1+r_\psi}{p_F+k^2\omega_\psi}\frac{\partial}{\partial q^2}\frac{1}{p_\varphi+k^2\omega_\varphi}\,,\\
m_{(12)R_\varphi}^{(FB)D,\sigma}(g_\varphi,g_\theta;\omega_\varphi,\omega_\theta,\omega_{\varphi\theta},\omega_\psi)&=-\frac{1}{4v_D}k^{4-D}\tilde{\partial}_t\int_q q^2\frac{1+r_\psi}{p_F+k^2\omega_\psi}\frac{\partial}{\partial q^2}\frac{g_\varphi(p_\theta+k^2\omega_\theta)-\sigma g_\theta k^2\omega_{\varphi\theta}}{(p_\varphi+k^2\omega_\varphi)(p_\theta+k^2\omega_\theta)-k^4\omega_{\varphi\theta}^2}\,.
\end{align*}
%


\subsection{Linear cutoff}

The regulator functions for the linear cutoff are explicitly given by 
\begin{align}
r_{\psi,k}(q)&=\rund{\frac{k}{q}-1}\theta(k^2-q^2)\,,&
r_{\chi/\phi,k}(q)&=\rund{\frac{k^2}{q^2}-1}\theta(k^2-q^2)\,,
\end{align}
with the step function $\theta(x)=1$ if $x>0$ and $\theta(x)=0$ if $x\leq0$.
Then, the threshold functions for the linear cutoff relevant to the flow equation for the effective potential, Eq.~\eqref{eq:potflow}, read
\begin{eqnarray}
	I_{R}(\omega_\chi,\omega_\phi,\omega_{\phi\chi}^2)&=&\frac{4 v_D}{D}\frac{(1-\frac{\eta_\phi}{D+2})(1+\omega_\chi)+(1-\frac{\eta_\chi}{D+2})(1+\omega_\phi)}{(1+\omega_\phi)(1+\omega_\chi)-\omega_{\phi\chi}^2}\,,\\
	I_{G}(\omega)&=&\frac{4 v_D}{D}\Big(1-\frac{\eta_\phi}{D+2}\Big)\frac{1}{1+\omega}\,,\\
	I_{\psi}(\omega)&=&\frac{4 v_D}{D}\Big(1-\frac{\eta_\psi}{D+1}\Big)\frac{1}{1+\omega}\,.
\end{eqnarray}
For the contributions to the Yukawa couplings we obtain the following threshold functions when evaluated with the linear cutoff
\begin{align*}
l_{(nm)R_{\varphi}}^{(FB),\sigma}(g_\varphi,g_\theta;\omega_\varphi,\omega_\theta,\omega_{\varphi\theta},\omega_\psi^\sigma)&=\frac{2}{D}\left[
-m\left(1-\frac{\eta_\theta}{D+2}\right)\frac{\sigma\omega_{\varphi\theta}(g_\theta(1+\omega_\varphi)-\sigma g_\varphi\omega_{\varphi\theta})}{(g_\varphi(1+\omega_\theta)-\sigma g_\theta\omega_{\varphi\theta})((1+\omega_\varphi)(1+\omega_\theta)-\omega_{\varphi\theta}^2)}\right.\\
&\left. \quad
+m\left(1-\frac{\eta_\varphi}{D+2}\right)\frac{(1+\omega_\theta)}{(1+\omega_\varphi)(1+\omega_\theta)-\omega_{\varphi\theta}^2}\right.\\
&\left. \quad+n\left(1-\frac{\eta_\psi}{D+1}\right)\frac{1}{1+\omega_\psi^{\sigma}}
\right]\frac{(g_\varphi(1+\omega_\theta)+\sigma g_\theta\omega_{\varphi\theta})^m}{\left((1+\omega_\varphi)(1+\omega_\theta)-\omega_{\varphi\theta}^2\right)^m(1+\omega_\psi^{\sigma})^n}\,,\\
l_{(nm)G_\phi}^{(FB)}(\omega_\phi,\omega_\psi)&=\frac{2}{D}\left[m\left(1-\frac{\eta_\phi}{D+2}\right)\frac{1}{1+\omega_\phi}+n\left(1-\frac{\eta_\psi}{D+1}\right)\frac{1}{1+\omega_\psi}\right]\frac{1}{(1+\omega_\phi)^m(1+\omega_\psi)^n}\,.
\end{align*}
and
\begin{align*}
l_{(111)R_{\varphi}R_{\theta}}^{(FBB),\sigma}(g_{\varphi},g_{\theta};\omega_{\varphi},\omega_{\theta},\omega_{\varphi\theta},\omega_\psi^\sigma)&=\frac{2}{D}\left[\left(1-\frac{\eta_\psi}{D+1}\right)\frac{(g_\varphi(1+\omega_\theta)-\sigma g_\theta\omega_{\varphi\theta})(g_\theta(1+\omega_\varphi)-\sigma g_\varphi\omega_{\varphi\theta})}{1+\omega_\psi^\sigma}\right.\\
+\left(1-\frac{\eta_\theta}{D+2}\right)&\frac{(1+\omega_\varphi)(g_\theta(1+\omega_\varphi)-\sigma g_\varphi\omega_{\varphi\theta})(g_\varphi(1+\omega_\theta)-\sigma g_\theta\omega_{\varphi\theta})-\sigma\omega_{\varphi\theta}(g_\theta(1+\omega_\varphi)-\sigma g_\varphi\omega_{\varphi\theta})^2}{(1+\omega_\varphi)(1+\omega_\theta)-\omega_{\varphi\theta}^2}\\
+\left(1-\frac{\eta_\varphi}{D+2}\right)&\left.\frac{(1+\omega_\theta)(g_\theta(1+\omega_\varphi)-\sigma g_\varphi\omega_{\varphi\theta})(g_\varphi(1+\omega_\theta)-\sigma g_\theta\omega_{\varphi\theta})-\sigma\omega_{\varphi\theta}(g_\varphi(1+\omega_\theta)-\sigma g_\varphi\omega_{\varphi\theta})^2}{(1+\omega_\varphi)(1+\omega_\theta)-\omega_{\varphi\theta}^2}\right]\\
&\times\frac{1}{((1+\omega_\varphi)(1+\omega_2)-\omega_{\varphi\theta}^2)^2(1+\omega_\psi^\sigma)}\,,\\
\end{align*}

\begin{align*}
l_{(111)G_{\phi}R_{\varphi}}^{(FBB),\sigma}(g_{\varphi},g_{\theta};\omega_\phi,\omega_{\varphi},\omega_{\theta},\omega_{\varphi\theta},\omega_\psi^\sigma)&=\frac{2}{D}\left[(1+\omega_\theta)\left(1-\frac{\eta_\varphi}{D+2}\right)\frac{g_\varphi(1+\omega_\theta)-\sigma g_\theta \omega_{\varphi\theta}}{(1+\omega_\varphi)(1+\omega_\theta)-\omega_{\varphi\theta}^2}\right.\\
&-\sigma\omega_{\varphi\theta}\left(1-\frac{\eta_\theta}{D+2}\right)\frac{g_\theta(1+\omega_\varphi)-\sigma g_\varphi\omega_{\varphi\theta}}{(1+\omega_\varphi)(1+\omega_\theta)-\omega_{\varphi\theta}^2}\\
&\left.+\left(1-\frac{\eta_\phi}{D+2}\right)\frac{g_\varphi(1+\omega_\theta)-\sigma g_\theta \omega_{\varphi\theta}}{1+\omega_\phi}+\left(1-\frac{\eta_\psi}{D+2}\right)\frac{g_\varphi(1+\omega_\theta)-
\sigma g_\theta \omega_{\varphi\theta}}{1+\omega_\psi^\sigma}\right]\\
&\times \frac{1}{(1+\omega_\phi)(1+\omega_\psi^\sigma)((1+\omega_\varphi)(1+\omega_\theta)-\omega_{\varphi\theta}^2)}\,,
\end{align*}
\begin{align*}
l_{(111)R_{\varphi}}^{(FFB),\sigma}(g_{\varphi},g_{\theta};\omega_{\varphi},\omega_{\theta},\omega_{\varphi\theta},\omega_{\psi1},\omega_{\psi2})=&\frac{2}{D}\left[(1+\omega_\theta)\left(1-\frac{\eta_\varphi}{D+2}\right)\frac{g_\varphi(1+\omega_\theta)-\sigma g_\theta\omega_{\varphi\theta}}{(1+\omega_\varphi)(1+\omega_\theta)-\sigma\omega_{\varphi\theta}^2}\right.\\
&-\sigma \omega_{\varphi\theta}\left(1-\frac{\eta_\theta}{D+2}\right)\frac{g_\theta(1+\omega_\varphi)-\sigma g_\varphi\omega_{\varphi\theta}}{(1+\omega_\varphi)(1+\omega_\theta)-\omega_{\varphi\theta}^2}\\
&+\left. \left(1-\frac{\eta_\psi}{D+1}\right)(g_\varphi(1+\omega_\theta)-\sigma g_\theta\omega_{\varphi
\theta})\left(\frac{1}{1+\omega_{\psi 1}}+\frac{1}{1+\omega_{\psi 2}}\right)\right]\\
&\times\frac{1}{(1+\omega_{\psi1})(1+\omega_{\psi2})((1+\omega_\varphi)(1+\omega_\theta)-\omega_{\varphi\theta}^2)}\,.
\end{align*}
The threshold functions for linear cutoff contributing with pure fermion loops to the anomalous dimensions are
\begin{align*}
m_4^{(F)D}(\omega)&=\frac{1}{(1+\omega)^4}+\frac{1-\eta_\psi}{D-2}\frac{1}{(1+\omega)^3}-\left(\frac{1-\eta_\psi}{2D-4}+\frac{1}{4}\right)\frac{1}{(1+\omega)^2}\,,\\
m_{(22)}^{(F)D}(\omega_1,\omega_2)&=\frac{1}{4}\left(\frac{1}{1+\omega_1}-\frac{2}{(1+\omega_1)^2}\right)\left(\frac{1}{1+\omega_2}-\frac{2}{(1+\omega_2)^2}\right)\\
&\quad+\frac{1}{4}\frac{1+\eta_\psi-D}{D-2}\left[\left(\frac{1}{1+\omega_1}-\frac{2}{(1+\omega_1)^2}\right)\frac{1}{1+\omega_2}+\left(\frac{1}{1+\omega_2}-\frac{2}{(1+\omega_2)^2}\right)\frac{1}{1+\omega_1}\right]\,,\\
m_2^{(F)D}(\omega)&=\frac{1}{(1+\omega)^4}\,,\\
m_{(11)}^{(F)D}(\omega)&=\frac{1}{(1+\omega_1)^2(1+\omega_2)^2}\,.
\end{align*}
The bosonic threshold functions are given by
\begin{align*}
m_{(22)R_\varphi R_\theta}^{(B)D}(v_1,v_2,v_3;\omega_\varphi,\omega_\theta,\omega_{\varphi\theta})=&(1+\omega_\varphi)(1+\omega_\theta)\frac{((1+\omega_\varphi)v_1-\omega_{\varphi\theta}v_2)((1+\omega_\theta)v_1-\omega_{\varphi\theta}v_3)}{((1+\omega_\varphi)(1+\omega_\theta)-\omega_{\varphi\theta})^4}\\
&+\omega_{\varphi\theta}^2 \frac{((1+\omega_\theta)v_2-\omega_{\varphi\theta}v_1)((1+\omega_\varphi)v_3-\omega_{\varphi\theta}v_1)}{((1+\omega_\varphi)(1+\omega_\theta)-\omega_{\varphi\theta})^4}\\
&-\omega_{\varphi\theta}(1+\omega_\theta)\frac{((1+\omega_\theta)v_2-\omega_{\varphi\theta}v_1)((1+\omega_\theta)v_1-\omega_{\varphi\theta}v_3)}{((1+\omega_\varphi)(1+\omega_\theta)-\omega_{\varphi\theta})^4}\\
&-\omega_{\varphi\theta}(1+\omega_\varphi)\frac{((1+\omega_\varphi)v_1-\omega_{\varphi\theta}v_2)((a+\omega_\varphi)v_3)-\omega_{\varphi\theta}v_1}{((1+\omega_\varphi)(1+\omega_\theta)-\omega_{\varphi\theta})^4}\,,\\
m_{(22)R_\varphi G_\phi}^{(B)D}(v_1,v_2;\omega_\varphi,\omega_\theta,\omega_{\varphi\theta},\omega_\phi)=&\frac{1+\omega_\theta}{(1+\omega_\phi)^2}\frac{(1+\omega_\theta)v_1-\omega_{\varphi\theta}v_2}{((1+\omega_\varphi)(1+\omega_\theta)-\omega_{\varphi\theta}^2)^2}-\frac{\omega_{\varphi\theta}}{(1+\omega_\phi)^2}\frac{(1+\omega_\chi)v_2-\omega_{\varphi\theta}v_1}{((1+\omega_\varphi)(1+\omega_\theta)-\omega_{\varphi\theta}^2)^2}\,,\\
m_{(40)R_\varphi}^{(B)D,\sigma}(v_1,v_2;\omega_\varphi,\omega_\theta,\omega_{\varphi\theta})=&\left[(1+\omega_\theta)\frac{(1+\omega_\theta)v_1-\omega_{\varphi\theta}v_2}{((1+\omega_\varphi)(1+\omega_\theta)-\omega_{\varphi\theta}^2)^2}-\omega_{\varphi\theta}\frac{(1+\omega_\theta)v_2-\omega_{\varphi\theta}v_1}{((1+\omega_\varphi)(1+\omega_\theta)-\omega_{\varphi\theta}^2)^2}\right]^2\,,\\
m_{(40)G_\phi}^{(B)D}(\omega_\phi)=&\frac{1}{(1+\omega_\phi)^4}\,.
\end{align*}
And finally, the mixed threshold functions read
\begin{align*}
m_{(12)G_\varphi}^{(FB)D}(\omega_{\varphi},\omega_\psi)&=\left(1-\frac{\eta_\varphi}{D+1}\right)\frac{1}{(1+\omega_\psi)(1+\omega_\varphi)^2}\,,\\
m_{(12)R_\varphi}^{(FB)D,\sigma}(g_\varphi,g_\theta;\omega_\varphi,\omega_\theta,\omega_{\varphi\theta},\omega_\psi)&=\frac{1}{1+\omega_\psi}\left[\left(1-\frac{\eta_\varphi}{D+1}\right)(1+\omega_\theta)\frac{g_\varphi(1+\omega_\theta)-\sigma g_\theta\omega_{\varphi\theta}}{((1+\omega_\varphi)(1+\omega_\theta)-\omega_{\varphi\theta}^2)^2}\right.\\
&\left.-\sigma\omega_{\varphi
\theta}\left(1-\frac{\eta_\theta}{D+1}\right)\frac{g_\theta(1+\omega_\varphi)-\sigma g_\varphi\omega_{\varphi\theta}}{((1+\omega_\varphi)(1+\omega_\theta)-\omega_{\varphi\theta}^2)^2}\right]\,.
\end{align*}
%


\subsection{Sharp cutoff}

We can use the threshold functions for the sharp cutoff to conveniently extract the $\epsilon$-expansion equations from the FRG $\beta$-functions. The sharp cutoff is given by~\cite{janssen2012}
\begin{align}
r_{\psi,k}(q)&=\lim_{a\rightarrow\infty}\rund{\sqrt{a\rund{\frac{k^2}{q^2}-\frac{a-1}{a}}}-1}\theta(k^2-q^2)\,, &
r_{\chi/\phi,k}(q)&=\lim_{a\rightarrow\infty}a\rund{\frac{k^2}{q^2}-1}\theta(k^2-q^2),
\end{align}
where the limit has to be taken after the integration over the loop momentum q. This gives the following explicit expressions. The contribution to the flow of the effective potential from the radial modes reads
\begin{align}
	I_{R}(\omega_\chi,\omega_\phi,\omega_{\phi\chi}^2)&=I_{R}^D(0,0,0)
	-2 v_D \log \Big((1+\omega_\phi)(1+\omega_\chi)-\omega_{\phi\chi}^2\Big)\,.
\end{align}
Accordingly, the expressions for the Goldstone and the fermionic contributions become
\begin{align}
	I_{G}(\omega)&=-2 v_D \log (1+\omega)+I_{G}^D(0)\,,\quad\text{and}\quad
	I_{\psi}(\omega)=-2 v_D \log (1+\omega)+I_{\psi}^D(0)\,.
\end{align}
Since we want to extract the $\epsilon$-expansion limit, we have to work in the SYM-SYM regime, which yields a large simplification in the $\beta$ functions for the Yukawa couplings and the anomalous dimensions. For the flow equations of the Yukawa couplings, we only need the threshold function
\begin{align}
\frac{1}{g_\varphi^m}l_{(nm)R_{\varphi}}^{(FB),\sigma}(g_\varphi,g_\theta;\omega_\varphi,\omega_\theta,\omega_{\varphi\theta},\omega_\psi^\sigma)\xrightarrow{\text{\scriptsize SYM}} l_{(nm)G_\phi}^{(FB)}(\omega_\varphi,\omega_\psi)\xrightarrow{\text{\scriptsize SYM}} l_{nm}^{(FB)}(\omega_\varphi,\omega_\psi)=\frac{1}{(1+\omega_\psi)^n(1+\omega_\varphi)^m}.
\end{align}
For the anomalous dimensions, we use
\begin{align}
m_{(22)}^{(F)D}(\omega_1,\omega_2)&\xrightarrow{\text{\scriptsize SYM}} m_4^{(F)}(\omega)=\frac{1}{(1+\omega)^4}\,,\\
\frac{1}{g_\varphi}m_{(12)R_\varphi}^{(FB)D,\sigma}(g_\varphi,g_\theta;\omega_\varphi,\omega_\theta,\omega_{\varphi\theta},\omega_\psi)&\xrightarrow{\text{\scriptsize SYM}}m_{(12)G_\varphi}^{(FB)D}(\omega_{\varphi},\omega_\psi)=\frac{1}{(1+\omega_\psi)(1+\omega_\varphi)^2}\,.
\end{align}


\section{Check of convergence}\label{app:conv}

We have checked the convergence of our polynomial expansion of the effective potential by extending our truncation up to 12th order in both fields $\chi$ and ${\boldsymbol\phi}$. 
In most cases the critical exponents are sufficiently convergent already at LPA8' level, with the remaining truncation error from the polynomial expansion being much smaller than the uncertainty due to the neglected higher-derivative operators.
The largest deviations occur for the third-largest critical exponent $\theta_3$ of the large-$N_f$ fixed point, as displayed in Fig.~\ref{fig:conv}. 
Still, already from LPA8' to LPA12' only minor improvements appear and we do not expect stronger deviations at even higher orders.
\begin{figure}[h!]
\centering
 \includegraphics[width=0.5\columnwidth]{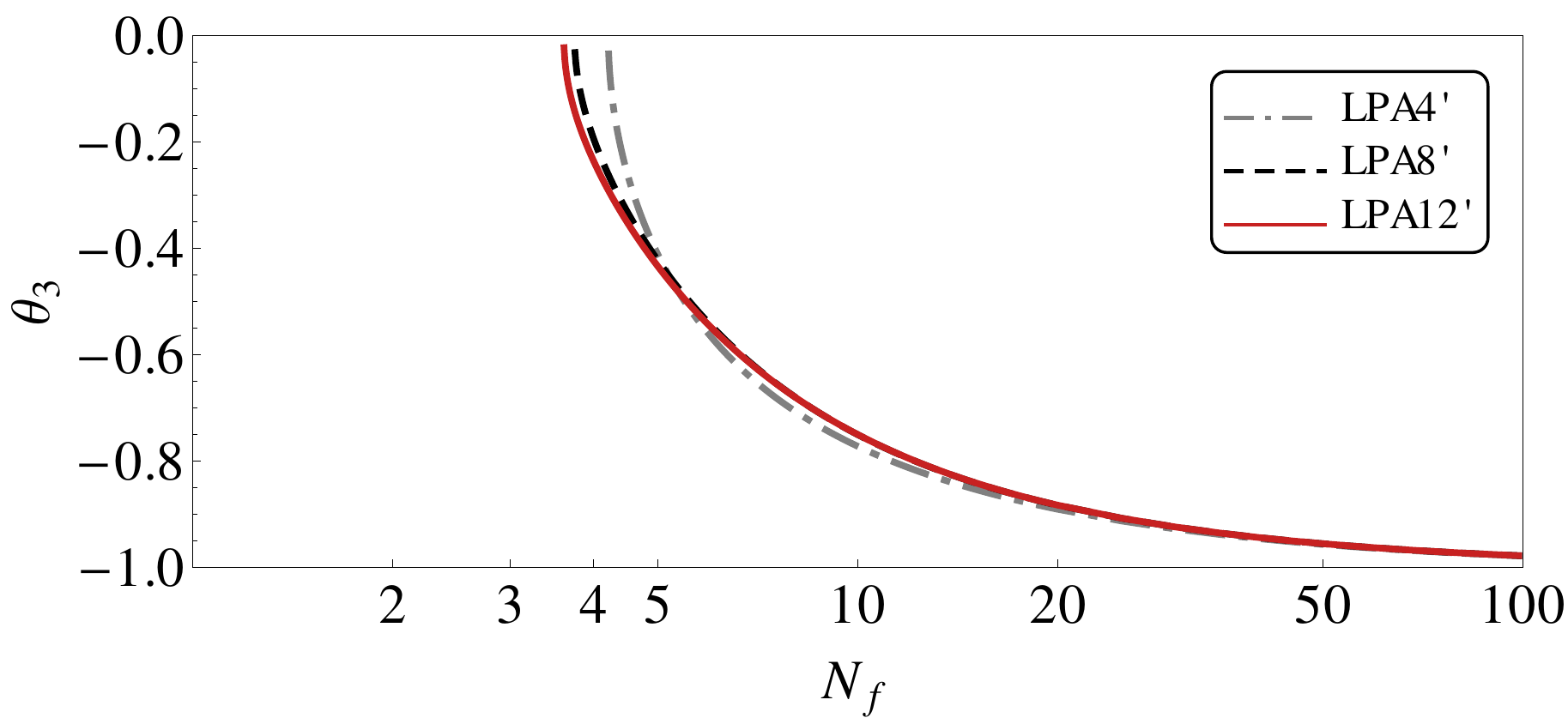}
 \caption{Third critical exponent in $D=3$ for large-$N_f$ fixed point as function of the fermion flavor number for different orders of polynomial truncation.}
\label{fig:conv}
\end{figure}

\end{widetext}


\end{document}